\RequirePackage{fix-cm}
\documentclass[smallextended]{svjour3}     

\smartqed  
\usepackage{appendix}
\usepackage{amsmath}
\usepackage{graphicx}
\usepackage{lineno}
\usepackage{array}
\usepackage{longtable}
\usepackage{natbib}
\linenumbers
\usepackage[a4paper,
left=3.2cm,
right=3.2cm,
top=3cm,
bottom=4cm]{geometry}

\newcommand*\patchAmsMathEnvironmentForLineno[1]{%
\expandafter\let\csname old#1\expandafter\endcsname\csname #1\endcsname
\expandafter\let\csname oldend#1\expandafter\endcsname\csname end#1\endcsname
\renewenvironment{#1}%
{\linenomath\csname old#1\endcsname}%
{\csname oldend#1\endcsname\endlinenomath}}%
\newcommand*\patchBothAmsMathEnvironmentsForLineno[1]{%
\patchAmsMathEnvironmentForLineno{#1}%
\patchAmsMathEnvironmentForLineno{#1*}}%
\AtBeginDocument{%
\patchBothAmsMathEnvironmentsForLineno{equation}%
\patchBothAmsMathEnvironmentsForLineno{align}%
\patchBothAmsMathEnvironmentsForLineno{flalign}%
\patchBothAmsMathEnvironmentsForLineno{alignat}%
\patchBothAmsMathEnvironmentsForLineno{gather}%
\patchBothAmsMathEnvironmentsForLineno{multline}%
}

\begin{document}
\nolinenumbers
\title{An improved fringe-region technique for the representation of gravity waves in large-eddy simulation with application to wind farms}
\titlerunning{An improved fringe-region technique for LES}       
\author{Luca Lanzilao \and Johan Meyers}
\institute{Luca Lanzilao \and Johan Meyers \at
           Department of Mechanical Engineering, KU Leuven, Celestijnenlaan 300A–Bus 2421, B3001  \\
           \email{luca.lanzilao@kuleuven.be}}
\maketitle

\begin{abstract}
	Large-eddy simulations of the atmospheric boundary layer are often performed using pseudo-spectral methods, which adopt a fringe-region approach to introduce inflow boundary conditions. However, we notice that a standard fringe-region technique excites spurious gravity waves when stratified atmospheres are considered, therefore enhancing the amount of energy reflected from the top of the domain and perturbing the velocity and pressure fields downstream. In this work, we develop a new fringe-region method that imposes the inflow conditions while limiting spurious effects on the surrounding flow. This is achieved by locally damping the convective term in the vertical momentum equation. We first apply the standard and wave-free fringe-region techniques to two-dimensional inviscid-flow simulations subjected to $169$ different atmospheric states. A similar study is performed on a three-dimensional domain using a couple of atmospheric states. In all cases, the new fringe-region technique outperforms the standard method, imposing the inflow conditions with a minimal impact on the surrounding flow. Moreover, we also investigate the performance of two already existing non-reflective upper boundary conditions, that is a Rayleigh damping layer (RDL) and a radiation condition (RC). Results highlight the importance of carefully tuning the RDL to limit the distortion of the numerical solution. Also, we find that the tuned RDL outperforms the RC in all cases. Finally, the tuned RDL together with the wave-free fringe-region method are applied to an LES of a wind farm operating in a conventionally neutral boundary layer, for which we measure a reflectivity of only $0.75\%$.
	\keywords{Fringe-region technique \and Gravity waves \and Large-eddy simulation \and Wind-farm simulation}
\end{abstract}

\section{Introduction}\label{sec--intro}
The constant increase of computational resources has made large-eddy simulation (LES) studies one of the most popular tools for analyzing the atmospheric boundary layer (ABL) responses to wind-farm forcing \citep{PorteAgel2020}. One of the key aspects that influences such dynamics is the thermal stratification. Offshore, the ABL is often characterized by a neutral boundary layer capped by an inversion layer and a stably stratified atmosphere aloft \citep{Smedman1997}. This type of ABL is often defined as a conventionally neutral boundary layer (CNBL) \citep{AllaertsPhD}. In such conditions, the momentum sink generated by the farm in the lower part of the ABL pushes upward the capping inversion, which in turns triggers gravity waves that propagate energy through the free atmosphere. To date, this phenomenon has been only investigated numerically, often with pseudo-spectral solvers which adopt a fringe-region approach to introduce inflow boundary conditions \citep{Stevens2014,Munters2016,Allaerts2017,Allaerts2017b,Wu2017,Gadde2021,Stieren2021}. This technique adds a body force to the right-hand side of the momentum and temperature equations, penalizing the error between the actual flow field and the desired inflow in a region usually located at the end of the domain \citep{Spalart1993,Lundbladh1999,Nordstrom1999,Inoue2014}. The desired inflow can be a simple unidirectional and laminar flow, or a fully developed turbulent velocity profile obtained with synthetic turbulence generators or with a concurrent precursor method \citep{Stevens2014,Dhamankar2015,Munters2016b}. As mentioned by \cite{Dhamankar2015}, an ideal inflow boundary condition should not introduce any spurious artefacts in the numerical solution. However, the capping inversion height at the end of the domain differs from the one of the inflow condition since the farm forcing displaces upward the inversion layer and triggers trapped gravity waves \citep{Nappo2002,Allaerts2019}. Hence, the body force applied within the fringe region inevitably modify the capping inversion height to restore the inflow conditions, thereby exciting spurious gravity waves and perturbing the velocity and pressure fields downstream. These waves are numerical artefacts which distort the flow fields in the domain of interest and enhance the amount of energy reflected from the top of the domain. The results presented by \cite{Allaerts2017,Allaerts2017b} show the presence of such spurious gravity waves, although they are not discussed by the authors. A similar behaviour is also observed in the work of \cite{Wu2017}, which uses a different pseudo-spectral solver from the one adopted in this manuscript. 

Wind-farm induced gravity waves absorb energy within the ABL, transporting it at higher altitudes
\citep{Smith2010,Smith2022,Wu2017,Allaerts2017,Allaerts2017b,Allaerts2019,Lanzilao2021,Devesse2022,Maas2022}. The energy is then released when the waves break down \citep{Nappo2002,Sutherland2010}. In a real-case scenario (and in the absence of potential temperature gradient variations in the free atmosphere), the energy is only transported upward since the source is located at the ground level. However, the boundaries of the computational domain allow the waves to reflect back, introducing disturbances within the domain. To overcome this issue, \cite{Klemp1977} introduced a Rayleigh damping layer (RDL) at the top of the domain to damp out gravity waves before they would have reached the upper boundary. Since then, this technique has been used extensively in mountain-wave simulations \citep{Klemp1977,Durran1983,Teixeira2014}, mesoscale models \citep{Klemp2008,Powers2017} and LES of large wind farms \citep{Allaerts2017,Wu2017,Allaerts2017b,Gadde2021,Maas2022}. The efficiency of this sponge layer increases with its vertical dimension. \cite{Klemp1977} recommended that the vertical extent should be about $3/2$ times the gravity-wave vertical wavelength, which ranges approximately between $1$ and $10$ km. Therefore, the RDL usually occupies a large part of the computational domain and as a result can be computationally expensive. This shortcoming was fixed by \cite{Beland1975} and \cite{Bennett1976}, who proposed an upper boundary condition which allows to radiate energy out of the domain without the need of a sponge layer. However, their technique requires evaluation of Laplace transforms, leading to unacceptable storage requirements when time integration is performed over a large number of time steps. \cite{Klemp1982} and \cite{Bougeault1982} were the first ones to propose an exact and parameter-free radiation boundary condition for linear hydrostatic Boussinesq equations in the absence of the Coriolis force, which did not require the computation of Laplace transforms. Despite the numerous assumptions, this radiation condition (RC) has been successfully used in several mesoscale models \citep{Jiang2004,Doyle2005,Klemp2008}. Finally, other studies have opted for both strategies, that is using a RDL in combination with a RC at the top of the domain \citep{Taylor2007,Taylor2008}. We note that also a perfectly matched layer technique could be adopted to avoid wave reflection \citep{Hu2008a,Hu2009}. However, when this method is applied to Navier--Stokes equations, it results in a coupled system of more than twenty absorbing boundary equations, becoming too complex and computationally expensive \citep{Hu2008b}. Therefore, we do not further consider this technique in the current manuscript. 

The goal of this work is twofold. First, we develop a wave-free fringe-region technique by locally damping the convective term in the vertical momentum equation to limit the advection of fringe-induced gravity waves into the domain of interest. Second, we show the importance of properly tuning the RDL to minimize the amount of energy reflected from the top of the domain. Moreover, we maximize its computational efficiency and we compare its performances against a RC. 

The article is further organized as follows. The numerical aspects together with the upper boundary conditions and fringe-region techniques are described in Sect.~\ref{sec--methodology}. To tune and test the standard and wave-free fringe-region methods, the RDL and the RC, we use two- and three-dimensional inviscid-flow simulations. These results are discussed in Sects. \ref{sec--2Dinviscidflow} and \ref{sec--3Dinviscidflow}, respectively. Subsequently, the methods are demonstrated in a LES of a wind farm in Sect. \ref{sec--LESresults}. Finally, conclusions are summarized in Sect.~\ref{sec--conclusions}.

\section{Methodology}\label{sec--methodology}

\subsection{Governing Equations}\label{sec--LESmodel}
The simulations performed in this study are based on the incompressible filtered Navier--Stokes equations coupled with a transport equation for the potential temperature and read as
\begin{align}
	& \frac{\partial \tilde{u}_i}{\partial x_i} = 0, \label{eq--continuity} \\
	& \frac{\partial \tilde{u}_i}{\partial t} + \frac{\partial}{\partial x_j} \bigl(\tilde{u}_j \tilde{u}_i \bigl) =  2 f_c \epsilon_{ij3} \tilde{u}_j + \delta_{i3} g \frac{\tilde{\theta}-\theta_0}{\theta_0} - \frac{\partial \tau_{ij}^{sgs}}{\partial x_j} - \frac{1}{\rho_0} \frac{\partial \tilde{p}^\ast}{\partial x_i} - \frac{1}{\rho_0} \frac{\partial p_\infty}{\partial x_i} + f_i^{tot}, \label{eq--momentum} \\
	& \frac{\partial \tilde{\theta}}{\partial t} + \frac{\partial }{\partial x_j} \bigl( \tilde{u}_j \tilde{\theta} \bigl) = - \frac{\partial q_j^{sgs}}{\partial x_j}, \label{eq--thermodynamic}
\end{align}
where the horizontal directions are denoted with $i=1,2$ while the vertical one is indicated by $i=3$. The filtered velocity and potential temperature fields are noted with $\tilde{u}_i$ and $\tilde{\theta}$. The first term on the right-hand side represents the Coriolis force due to planetary rotation, where the frequency $f_c$ depends on the Earth's latitude. Thermal buoyancy is taken into account by the second term, where $g=9.81$ m s\textsuperscript{-2} denotes the gravitational constant and $\theta_0$ a reference temperature. Several orders of magnitude separate the smallest and largest scales in boundary-layer flows, therefore we omit the resolved effects of viscous momentum transfer and diffusive heat transfer. Instead, these effects are modelled by the subgrid-scale stress tensor $\tau_{ij}^{sgs}$ and the subgrid-scale heat flux $q_j^{sgs}$. The filtered modified pressure, denoted with $\tilde{p}^\ast$, represents pressure fluctuations around a steady background pressure $p_\infty$, which is used to drive the flow across the domain. Note that the trace of the subgrid-scale stress tensor is absorbed into $\tilde{p}^\ast$. Finally, the term $f_i^{tot} = f_i + f_i^{ra} + f_i^{fr}$ groups all external forces exerted on the flow. Here, $f_i^{ra}$ and $f_i^{fr}$ represent the body forces applied within the RDL and fringe region, respectively, while $f_i$ denotes a generic force applied within the ABL, e.g. originating from wind turbines. The notation $(x_1,x_2,x_3)$ and $(x,y,z)$ together with $(\tilde{u}_1,\tilde{u}_2,\tilde{u}_3)$ and $(\tilde{u},\tilde{v},\tilde{w})$ are used interchangeably. Moreover, for sake of simplicity, the tilde will not be used in the rest of the manuscript.

The LES code SP-Wind is used to solve the governing equations \citep{Calaf2010,Goit2015,Allaerts2017,Munters2018,Allaerts2017b}. The solver uses a Fourier pseudo-spectral method to discretize the streamwise ($x$) and spanwise ($y$) directions while for the vertical dimension ($z$) a symmetry-preserving fourth-order finite difference scheme is adopted~\citep{Verstappen2003}. A classic fourth-order Runge--Kutta scheme is used for the temporal component, with a time step based on a Courant--Friedrichs--Lewy number of $0.4$. A Smagorinsky type model proposed by \cite{Stevens2000} is used to account for the effects of subgrid-scale motions on the resolved flow. The Smagorinsky coefficient is set to $C_s=0.14$, similarly to \cite{Calaf2010} and \cite{Allaerts2017}. The wall damping function used by \cite{Mason1992} is adopted to damp $C_s$ near the wall. The Poisson equation is solved during every stage of the Runge--Kutta scheme to enforce continuity. The constant pressure gradient which drives the flow through the domain is related to the geostrophic wind speed $G$ through the geostrophic balance. The effect of the bottom wall on the flow is modelled with classic Monin--Obukhov similarity theory for neutral boundary layers \citep{Moeng1984,Allaerts2017}. This wall-stress boundary condition is only dependent on the surface roughness $z_0$, which we assume to be constant. The upper boundary conditions together with the standard and wave-free fringe-region techniques are described in the next sections.

\subsection{Upper Boundary Condition}\label{sec--topBC}
Two different types of non-reflective upper boundary conditions are investigated in the current study, that is the RDL and the RC.

The RDL was initially introduced by \cite{Klemp1977} to damp out mountain-induced gravity waves at the top of the domain. It consists of an additional term on the right-hand side of the momentum equations which forces the flow to an unperturbed state, therefore dissipating the upward energy transport. This body force is applied within a sponge layer located at the top of the domain and reads as
\begin{equation*}
	f_i^{ra} (\boldsymbol{x}) = - \nu(z) \biggl( u_i(\boldsymbol{x}) - U_{G,i} \biggl)
\end{equation*}
where $U_{G,1} = G \cos{\alpha}$, $U_{G,2} = G \sin{\alpha}$ and $U_{G,3} = 0$ with $\alpha$ the geostrophic wind angle. The buffer layer performance depends on the Rayleigh function $\nu(z)$. This one-dimensional function should increase gradually with height to minimize wave reflection and it should be strong enough to dissipate upward-going energy. To this end, we choose
\begin{equation*}
	\nu(z) = \check{\nu} \biggl[1 - \cos{\biggl( \frac{\pi}{s^{ra}} \frac{z - \bigl(L_z-L_z^{ra} \bigl)}{L_z^{ra}}\biggl)} \biggl]
\end{equation*} 
where $\check{\nu}$ is an inverse decay time and controls the force magnitude while $s^{ra}$ regulates the function gradient along the vertical direction. Moreover, $L_z$ and $L_z^{ra}$ denote the height of the computational domain and of the RDL, respectively. \cite{Klemp1977} and \cite{Allaerts2017,Allaerts2017b} chose a similar Rayleigh function with $s^{ra}=~1$. Instead, we consider this as a free parameter. Moreover, we scale $\check{\nu}$ with the Brunt--V\"{a}is\"{a}l\"{a} frequency, i.e. $\check{\nu} = \nu^{ra}N$ with $N = \sqrt{g \Gamma/ \theta_0}$, where $\Gamma$ denotes the lapse rate in the free atmosphere. Hence, the RDL performance depends upon the choice of the non-dimensional parameters $\nu^{ra}$ and $s^{ra}$ while its computational efficiency is related to the number of vertical grid points spread over $L_z^{ra}$. The tuning of these parameters is performed in Sect. \ref{sec--2D_tuning_RDL}. Figure \ref{fig--sponge_layers_setup}a shows the Rayleigh function obtained with $L_z^{ra}=10$ km, $\nu^{ra}=3$ and different $s^{ra}$ values. We verified that if $\nu(z)$ is properly calibrated, the boundary condition specified at the top of the domain has limited influence on the numerical solution. Here, we impose zero shear and zero vertical velocity.

Next, we have implemented the RC proposed by \cite{Klemp1982} and \cite{Bougeault1982}. They were the first ones to propose an exact RC for the linear hydrostatic Boussinesq equations with a homogenous mean state in the absence of the Coriolis force, which reads as
\begin{equation}
	\frac{1}{\rho_0} \hat{p}(k,l,z_T,t) = \frac{N}{( k^2 + l^2)^{1/2}} \hat{w}(k,l,z_T,t)
	\label{eq--radiation_condition}
\end{equation}
where $\hat{p}$ and $\hat{w}$ represent the horizontal Fourier coefficients of the pressure and vertical velocity, respectively, taken at the top of the domain $z_T$. Note that $z_T = L_z - L_z^{ra}$ when the RC is used. Moreover, $k$ and $l$ denote the wavenumber in the $x$- and $y$-direction. Equation~\ref{eq--radiation_condition} shows that pressure and vertical velocity are related by a positive and time-independent quantity. The implication is twofold. First, a positive correlation between pressure and vertical velocity ensures a positive energy flux at the top of the domain. Second, the condition is local in time, avoiding the need of Laplace transforms. Despite the fact that the RC in Eq. \ref{eq--radiation_condition} has been derived for linear hydrostatic Boussinesq equations, \cite{Klemp1982} show that this upper boundary condition performs well even when these assumptions are violated. However, in such a case, Eq. \ref{eq--radiation_condition} is not exact anymore, therefore partial wave reflection may be expected. The use of a RC implies $f_i^{ra}=0$ and a computational domain height of $L_z - L_z^{ra}$ instead of $L_z$ in case of Rayleigh damping. 

Internal gravity waves have an intrinsic property so that the sign of the vertical phase velocity is opposite to the sign of the vertical group velocity. The method developed by \cite{Taylor2007} and later adopted by \cite{Allaerts2017,Allaerts2017b} makes use of this property to distinguish upward from downward going waves in a frequency domain. Next, an inverse Fourier transform yields the vertical velocity perturbations divided into upward and downward internal waves. Similarly to \cite{AllaertsPhD,Allaerts2017}, the corresponding vertical kinetic energy is then computed over a $x$-$z$ plane including the free atmosphere only without the buffer regions. The ratio between the vertical kinetic energy associated with downward and upward internal waves is what we define as reflectivity $r$. Since the only momentum sink which triggers gravity waves is located at the surface, internal waves transporting energy downward are solely due to reflection from the top of the domain. Therefore, the closer $r$ is to zero, the better the upper boundary condition performs.

\subsection{Fringe-Region Technique}\label{sec--newfringe}
A volume force applied within the fringe region is used to impose the desired inflow conditions, which we denote with $u_{\mathrm{in},i}(\boldsymbol{x})$. The fringe forcing term reads as
\begin{equation*}
	f_i^{fr}(\boldsymbol{x}) = -h(x) \biggl( u_i(\boldsymbol{x}) - u_{\mathrm{in},i}(\boldsymbol{x})\biggl)
\end{equation*}
where $h = h(x)$ is a one-dimensional non-negative function which is non-zero only within the fringe region, and is expressed as 
\begin{equation*}
	h(x) = h_\mathrm{max} \biggl[ F\biggl( \frac{x-x_s^h}{\delta_s^h}\biggl) -  F\biggl( \frac{x-x_e^h}{\delta_e^h} +1\biggl) \biggl]
\end{equation*}
with
\begin{equation*}
	F(x) = \begin{cases}
		0,& \text{if } x \leq 0\\
		\dfrac{1}{1+\text{exp}\biggl( \dfrac{1}{x-1} + \dfrac{1}{x}\biggl)},& \text{if } 0<x<1\\
		1,& \text{if } x \geq 1.
	\end{cases}
\end{equation*}
We choose this expression for the function $F(x)$ since it has the advantage of having continuous derivatives of all orders \citep{Lundbladh1999,Nordstrom1999}. The parameters $x_s^h$ and $x_e^h$ denote the start and end of the fringe function support while its smoothness is regulated by $\delta_s^h$ and $\delta_e^h$. Moreover, $h_\mathrm{max}$ denotes the maximum value of the fringe function. This technique, which we refer to as the standard fringe-region method, causes negligible disturbances in the surrounding flow when applied to pressure-driven boundary layer flows \citep{Nordstrom1999,Goit2015,Munters2016}. However, when applied to CNBL or stable boundary layer (SBL) flows, this body force displaces vertically air parcels with different temperatures while restoring the inflow condition. As a result, spurious gravity waves are triggered which propagate through the domain of interest, causing a distortion of the vertical velocity field which in turns perturbs the streamwise velocity and pressure fields.

To overcome this issue, we propose a new technique which imposes the desired inflow while locally damping the convective term in the vertical momentum equation. To this end, we define the following damping function 
\begin{equation}
	d(x,z) = 1 -  \biggl[ F\biggl( \frac{x-x_s^d}{\delta_s^d}\biggl) -  F\biggl( \frac{x-x_e^d}{\delta_e^d} +1\biggl) \biggl] \mathcal{H}(z-H)
	\label{eq--damping_function}
\end{equation}
where $x_s^d$ and $x_e^d$ define the start and end of the damping function support while $\delta_s^d$ and $\delta_e^d$ control the function smoothness. Moreover, $H$ denotes the capping-inversion base height (which also corresponds to the ABL height) while $\mathcal{H}$ represents a Heaviside function. We note that a height-dependent damping function is necessary in case of a turbulent inflow profile. In fact, if the damping would be applied also within the ABL, not only spurious gravity waves would be dampened but also turbulence fluctuations (see Sect.~\ref{sec--LESresults} for more details). The modified vertical momentum equation then corresponds to
\begin{equation*}
	\frac{\partial u_3}{\partial t} + d(x_1,x_3)\frac{\partial}{\partial x_j} \bigl(u_j u_3 \bigl) = g \frac{\theta-\theta_0}{\theta_0} - \frac{\partial \tau_{3j}^{sgs}}{\partial x_j} - \frac{1}{\rho_0} \frac{\partial p^\ast}{\partial x_3} - \frac{1}{\rho_0} \frac{\partial p_\infty}{\partial x_3} + f_3^{tot}.
\end{equation*}
Note that in boundary-layer flows the terms $\partial (u_2 u_3) / \partial x_2$ and $\partial (u_3 u_3) / \partial x_3$ have a smaller order of magnitude than the term $\partial (u_1 u_3) / \partial x_1$. Therefore, for the cases that we have examined in this work, we found that similar results are obtained if the local damping would be applied only to $\partial (u_1 u_3) / \partial x_1$ (results not further discussed here).  
\begin{figure}[t!]
	\centering
	\includegraphics[width=1\textwidth]{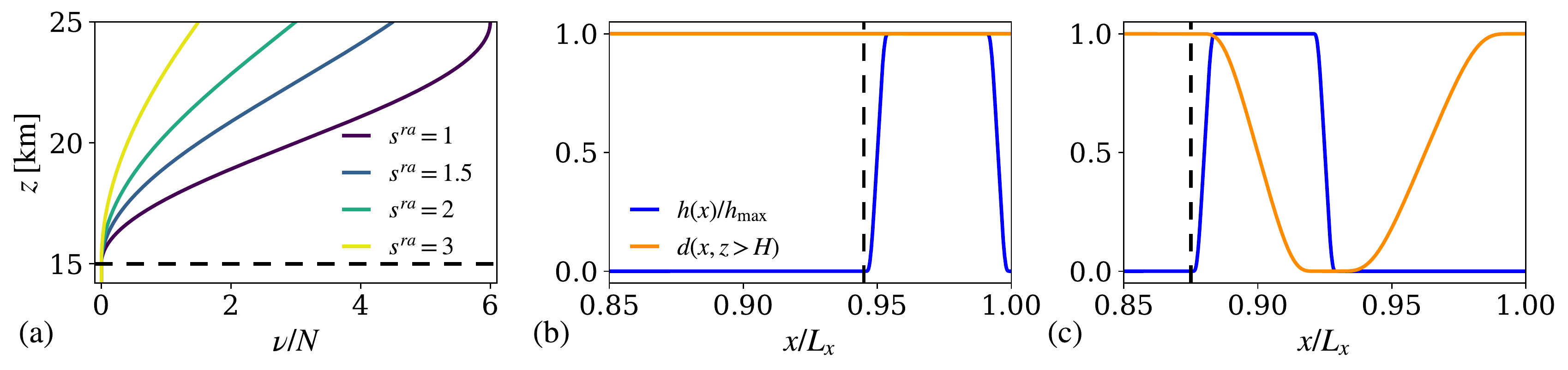}%
	\caption{(a) Rayleigh function obtained with $\nu^{ra}=3$ and different $s^{ra}$ values. (b) Fringe and damping functions used with the standard fringe-region technique. The fringe function parameters are set to $x_s^h=0.945 L_x$, $x_e^h=L_x$ and $\delta_s^h=\delta_e^h= 0.01 L_x$ while $d(x,z)=1$ everywhere. (c) Fringe and damping functions used with the wave-free fringe-region technique. The fringe function parameters are set to $x_s^h=0.875 L_x$, $x_e^h=0.93 L_x$ and $\delta_s^h=\delta_e^h= 0.01 L_x$ while the damping function ones are fixed to $x_s^d=x_s^h$, $x_e^d=L_x$, $\delta_s^d=0.05 L_x$ and $\delta_e^d= 0.075 L_x$. Note that $L_x$ denotes the streamwise domain length. Moreover, $h_\mathrm{max}=0.03$~s\textsuperscript{-1} as in \cite{Allaerts2017,Allaerts2017b}. Finally, the black horizontal and vertical dashed lines denote the start of the RDL and the fringe region}
	\label{fig--sponge_layers_setup}
\end{figure}

We did extensive testing on the fringe and damping-function setups (results not shown in detail), and found that the technique is effective when the damping function is zero in the region where the fringe function decreases from $h_\mathrm{max}$ to zero. Moreover, we notice that the new fringe-region technique performs well when enough grid points along the streamwise direction (at least $8$ according to our tests) are located in the region where $d(x,z)=0$, which is usually the case given the fine horizontal grid resolution of LES performed nowadays \citep{Allaerts2017,Allaerts2017b,Gadde2021,Lanzilao2022}. Furthermore, a smooth damping function in the $x$-direction is needed to avoid numerical oscillations. Finally, we note that the fringe forcing should be strong enough to force the desired inflow condition without violating the stability constraint imposed by the 4th order Runge--Kutta method, i.e. $h_\mathrm{max} \Delta t \leq 2.78$ with  $\Delta t$ denoting the time step \citep{Schlatter2005}. If these constraints are satisfied, the inflow profile is correctly imposed and the fringe-induced gravity waves remain trapped within the fringe region, avoiding the distortion of the velocity and pressure fields downstream. Figure \ref{fig--sponge_layers_setup}b, c shows the setup of the standard and wave-free fringe and damping functions, respectively, obtained following the guidelines defined above. The support, shape and maximum of the fringe function are equal in both cases. However, a longer buffer region is necessary with the new technique to allow room for a sufficiently smooth damping function. With the current setup, $12.5\%$ of the total grid cells in the main domain are located within the fringe region when the new technique is employed while this number reduces to $5.5\%$ for the standard method. The length of the fringe region, which we denote with $L_x^{fr}$, is determined as $x_e^h-x_s^h$ and $\max\{x_e^h,x_e^d\}-\min\{x_s^h,x_s^d\}$ in the standard and wave-free fringe-region technique, respectively. 

\section{Two-Dimensional Inviscid-Flow Simulations}\label{sec--2Dinviscidflow}
In a first step, we test the fringe-region techniques together with the non-reflective upper boundary conditions in a two-dimensional ($x$--$z$) inviscid-flow environment. Here, the use of the sub-grid scale model is not necessary. The absence of viscous forces allows us to perform simulations with a relatively coarse grid resolution. Moreover, the flow is not driven by a background pressure gradient. Instead, a unidirectional and constant with height inflow velocity profile (i.e. $u_{\mathrm{in},1}(\boldsymbol{x}) = U_\infty$ and $u_{\mathrm{in},2}(\boldsymbol{x}) = u_{\mathrm{in},3}(\boldsymbol{x}) = 0$ with $\alpha = 0 ^\circ$), which is imposed within the fringe region, drives the flow. To comply with the assumptions on the inflow velocity profile, the Coriolis force and the wall stress are not applied. Instead, a symmetry boundary condition is used at the bottom. The potential temperature profile consists of a neutral ABL capped by an inversion layer with height $H$, strength $\Delta \theta$ and depth $\Delta H$ and a free atmosphere aloft with a constant lapse rate~$\Gamma$. We use the model developed by \cite{Rampanelli2004} to define such a temperature profile, which characterizes a CNBL. Finally, a smooth box-like force model with length scale $L_x^s$ emulates the presence of a wind-farm, generating a momentum sink within the ABL (see Appendix 1 for more details). These choices make the flow solver several orders of magnitude faster than standard LES of wind farms, allowing us to easily explore the parameter spaces of interest. Note that this setup is inspired by the work of~\cite{AllaertsPhD}, Appendix C.

Next, we non-dimensionalize the governing equations using a recurring set of three variables, that is the inflow velocity $U_\infty$, the height of the capping inversion $H$ and the surface temperature $\theta_0$. This allows us to derive the following non-dimensional groups
\begin{equation}
	\pi_g = \frac{gH}{U_\infty^2}, \;\; \pi_{\Delta H} = \frac{\Delta H}{H} \;\; \pi_{\Delta \theta} = \frac{\Delta \theta}{\theta_0}, \;\; \pi_{\Gamma} = \frac{\Gamma H}{\theta_0}, \;\; \pi_L=\frac{L_x^s}{H}.
	\label{eq--non-dim_groups}
\end{equation}
In the remainder of this section, we fix $U_\infty=12$~m s\textsuperscript{-1}, $H=1000$~m, $\Delta H = 100$~m and $\theta_0=288.15$ K, which define $\pi_g=68.125$ and $\pi_{\Delta H} =0.1$. Next, we will first select a set of $\pi_{\Delta \theta}$, $\pi_\Gamma$ and $\pi_L$ values for tuning the RDL in Sect. \ref{sec--2D_tuning_RDL}. We will use these cases to analyze in detail the flow response to the momentum sink applied within the ABL when four different numerical setups are used in Sect.~\ref{sec--2D_flow_physics}. Finally, we perform a sensitivity analysis of several quantities of interest by extensively varying $\pi_{\Delta \theta}$, $\pi_{\Gamma}$ and $\pi_L$ in Sect.~\ref{sec--2D_sensitivity_study}. We note that the Froude ($\textit{Fr}$) and $P_N$ numbers can be written as $\textit{Fr}= 1/\sqrt{\pi_g \pi_{\Delta \theta}}$ and $P_N = 1/\sqrt{\pi_g \pi_{\Gamma}}$, meaning that we will effectively vary the two non-dimensional quantities that characterize the impact of gravity waves on the flow dynamics \citep{Smith2010,Allaerts2017b,Allaerts2019,Lanzilao2021}.

\subsection{Tuning of the RDL Parameters}\label{sec--2D_tuning_RDL}
In this section, we tune the RDL by performing approximately thousand simulations sweeping through the $\nu^{ra}$--$s^{ra}$ parameter space in search of the values that minimize the reflectivity. To do so, we use the atmospheric states and box-like forcing region lengths considered by \cite{AllaertsPhD}.  Hence, we set $\pi_{\Delta H}=0.1$ and $\pi_{\Delta \theta}=3.47\times 10^{-3}$ which correspond to a capping inversion $100$ m deep with a strength of $1$ K. For the free atmosphere, we use two values of $\pi_{\Gamma}$, that is $3.47\times 10^{-3}$ and $3.47\times 10^{-2}$, which corresponds to a weakly and strongly stratified atmosphere. Further, we fix two values of $\pi_L$, i.e. $5$ and $15$. Note that the force integrated over the whole computational domain has equal magnitude for the two cases (see Appendix 1), which therefore emulate the presence of a small dense ($\pi_L=5$) and a large sparse ($\pi_L=15$) wind farm. For the computational domain, we fix the length and height to $40$ km and $25$ km, which is in accordance with previous studies \citep{Allaerts2017,Allaerts2017b}. In the horizontal direction we use a uniform grid with $N_x=256$ grid points, while in the vertical direction a stretched grid is used. The latter is composed of $300$ uniformly spaced grid points within the first $1.5$ km. A first stretch is applied from $1.5$ km to $15$ km, where $180$ points are used. A second one is applied within the RDL, i.e from $15$ km to $25$ km (i.e. $L_z^{ra}=10$ km). We have tested several grid stretchings within the RDL spanning from $100$ to $5$ grid points and we have noticed that $10$ grid points suffice to damp out waves before they reach the top of the domain (not shown). However, we note that we did not perform sensitivity tests on the height of the RDL. Instead, we used the same $L_z^{ra}$ adopted in previous studies \citep{Allaerts2017,Allaerts2017b}. These choices make the RDL extremely efficient, since only the $2$\% of vertical grid points are located in this region. To summarize, the vertical grid contains a total of $490$ grid points. 
\begin{figure}[t!]
	\centering
	\includegraphics[width=1\textwidth]{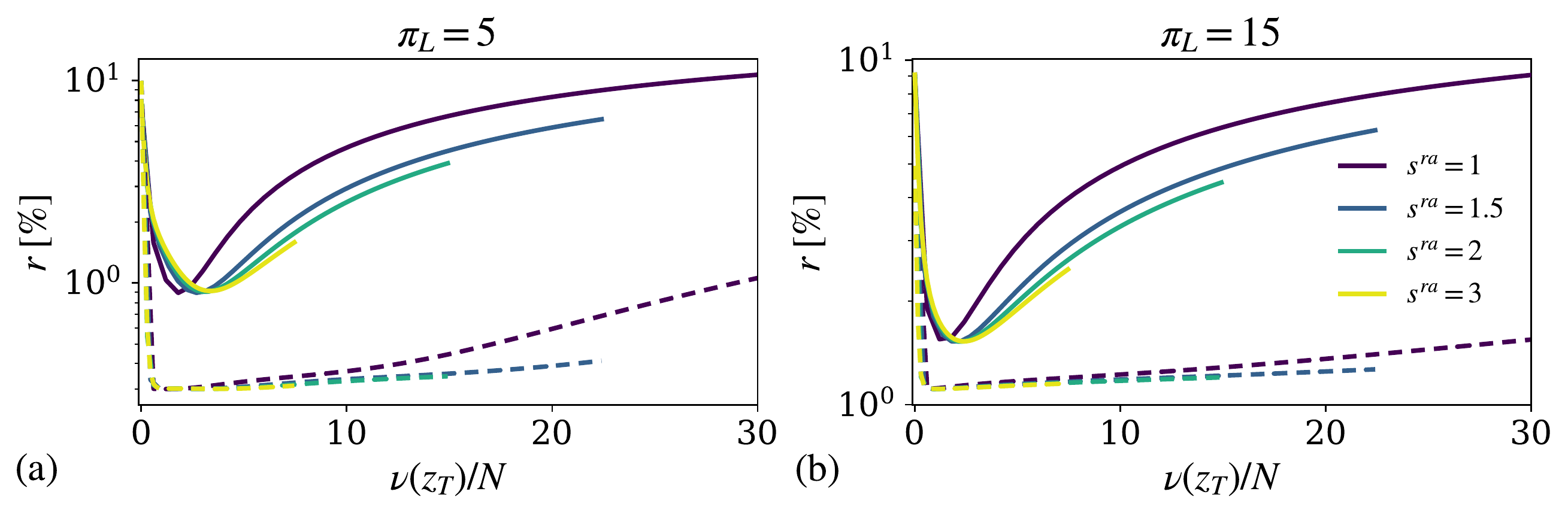}%
	\caption{Reflectivity as function of $\nu(z_T)/N$ (i.e. the maximum Rayleigh function value normalized with the Brunt--V\"{a}is\"{a}l\"{a} frequency) obtained with (a) $\pi_L=5$ and (b) $\pi_L=15$. The parameter $\nu^{ra}$ is varied between $0$ and $15$ while $s^{ra}$ varies between $1$ and $3$. The solid and dashed lines refer to results obtained with $\pi_{\Gamma}=3.47\times 10^{-3}$ (weakly stratified atmosphere) and $\pi_{\Gamma}=3.47\times 10^{-2}$ (strongly stratified atmosphere), respectively. Note that the $y$-axis is in logarithmic scale}
	\label{fig--RDL_tuning}
\end{figure}

The reflectivity as function of $\nu(z_T)/N$ (i.e. the maximum Rayleigh function value normalized with the Brunt--V\"{a}is\"{a}l\"{a} frequency - see Fig. \ref{fig--sponge_layers_setup}a) is shown in Fig. \ref{fig--RDL_tuning}. Here, we vary $\nu^{ra}$ between $0$ and $15$ and $s^{ra}$ between $1$ and $3$. Setting $\nu^{ra}$ to zero corresponds to a case without RDL, for which the reflectivity is about $10\%$ in all cases. As $\nu^{ra}$ increases, the reflectivity reaches a global minimum, increasing monotonically afterwards. Such a minimum is almost insensitive to the choice of $s^{ra}$. Similar trends have been observed by \cite{Klemp1977}. For both $\pi_L$ values, the reflectivity is low and rather constant to changes in the Rayleigh function when $\pi_{\Gamma}=3.47\times 10^{-2}$ (strongly stratified atmosphere). This is explained by the fact that $L_z^{ra}$ is approximately $5/2$ times the gravity-wave vertical wavelength $\lambda_z$, which is evaluated as $2 \pi U_\infty/N$. Conversely, a higher sensitivity is observed when $\pi_{\Gamma}=3.47\times 10^{-3}$ (weakly stratified atmosphere), where the $L_z^{ra}$ to $\lambda_z$ ratio is roughly one. Figure \ref{fig--RDL_tuning} also shows that a non-tuned RDL could lead to a higher reflectivity than a setup without RDL, highlighting the importance of properly calibrating the Rayleigh function. The minimum reflectivity value in all atmospheric states is attained with the parameters $\nu^{ra}=3$ and $s^{ra}=2$, which we use to tune the RDL. The Rayleigh function obtained with these parameters is shown in Fig. \ref{fig--sponge_layers_setup}a. In contrast to the RDL, the RC is parameter-free, therefore there is no need to perform a calibration study. Moreover, the vertical grid contains $480$ points instead of 490 when the RC is adopted. 

\subsection{Flow Physics}\label{sec--2D_flow_physics}
To show the differences in the numerical solution between a standard upper boundary condition and a non-reflective one, we use the standard fringe-region technique without and with the tuned RDL and we name these cases as stdFR-woRDL and stdFR-RDL, respectively. Next, we adopt the new fringe-region technique together with the tuned RDL (newFR-RDL). The comparison between cases stdFR-RDL and newFR-RDL will highlight the benefits of using a wave-free fringe-region technique. Finally, we use the new fringe-region method together with the RC (newFR-RC). The newFR-RDL and newFR-RC cases allow us to asses the performances of the two different non-reflective upper boundary conditions. Since adopting the standard fringe-region technique with the RC would not bring further insights, we decided to not consider this case. Hence, we have a total of four different numerical setups which we drive with the same atmospheric states and box-like force term that were used in Sect. \ref{sec--2D_tuning_RDL}.

The simulations are performed on a domain with length $L_x = 40$ km, as in Sect. \ref{sec--2D_tuning_RDL}. However, we vary $N_x$ from $256$ to $1000$, leading to a horizontal grid resolution of $40$ m. The grid sensitivity study performed in Appendix 2 shows that this change has almost no impact on the reflectivity value, and that the numerical solution is grid independent. Finally, we define two simulations which are performed on a domain five times longer (i.e. $200$~km long) with an equal grid resolution. Both simulations adopt the tuned RDL, but one makes use of the standard fringe region-technique (stdFR-RDL-LD) while the second employs the new technique (newFR-RDL-LD). We show in Appendix 3 that the length of the domain is large enough for the newFR-RDL-LD solution to be not affected by the boundary conditions. Consequently, the numerical solutions obtained on the small domain of $40$ km will be compared against the newFR-RDL-LD solution in the remainder of this section. For this reason, the latter case will be also named as the reference (Reference). In all cases, the time horizon is set to $T=2$ h. At this point, a stationary gravity-wave pattern has formed and the flow has reached a steady state \citep{AllaertsPhD}. 
\begin{figure}[t!]
	\centering
	\includegraphics[width=1\textwidth]{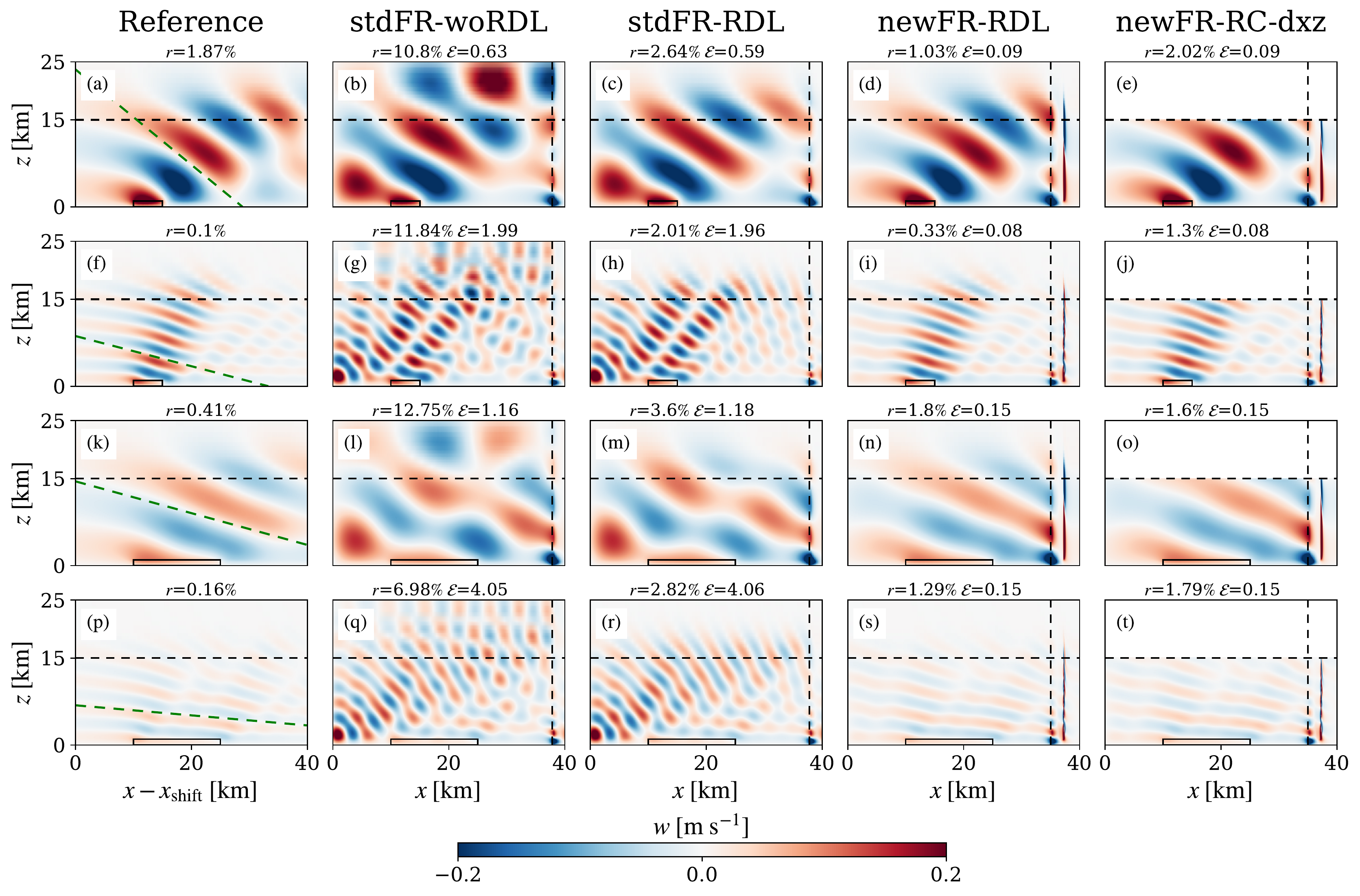}%
	\caption{Side view of vertical velocity obtained with (a-e) $\pi_L=5$ and $\pi_\Gamma=3.47\times 10^{-3}$, (f-j) $\pi_L=5$ and $\pi_\Gamma=3.47\times 10^{-2}$, (k-o) $\pi_L=15$ and $\pi_\Gamma=3.47\times 10^{-3}$ and (p-t) $\pi_L=15$ and $\pi_\Gamma=3.47\times 10^{-2}$. Note that the solution in the reference domain is shifted of $x_\mathrm{shift}=90$~km. The black solid lines denote the region in which the box-like force term is applied. The black horizontal and vertical dashed lines indicate the starting location of the RDL and the fringe region. Finally, the green dash-dotted line in (a,f,k,p) denotes the inclination of the gravity-wave phase line predicted by linear theory}
	\label{fig--2D_w}
\end{figure}

Gravity waves are triggered by the box-like force term applied within the ABL and propagate through the free atmosphere. Figure \ref{fig--2D_w}a, f, k, p shows the vertical velocity contours obtained with the reference setup together with a dash-dotted green line which represents the angle that the gravity-wave phase lines make with the horizontal. This angle is calculated using linear theory as $\arccos{ \bigl( k_x U_\infty/N\bigl)}$ with $k_x=1/L_x^s$ \citep{Lin2007}. Good agreement between reference simulations and linear theory is found for all $\pi_\Gamma$ and $\pi_L$ values. Moreover, the highest vertical velocity value is found near the leading edge of the box-like forcing region, as a response to the momentum sink applied within the ABL. Next, Fig.~\ref{fig--2D_w}b, g, l, q displays the results obtained with the stdFR-woRDL setup. Here, it is evident how the fringe body force also triggers spurious gravity waves which heavily perturb the flow field downstream, altering the gravity-wave phase lines. This is particularly noticeable when a strong free lapse rate is used. The same phenomenon is observed in the stdFR-RDL case, where the maximum $w$ value is attained near the inflow region instead of at the leading edge of the box-like forcing region. Conversely, the new fringe-region technique traps the spurious gravity waves within the fringe region by damping the convective term in the vertical momentum equation. This is visible in the newFR-RDL and newFR-RC cases. As a result, the numerical solution within and around the box-like forcing region is very similar to the reference one, despite using a domain which is five times smaller in length. 

In terms of reflectivity, high values are obtained when a RDL is not used. However, results show that the reflectivity can be reduced by one order of magnitude when the RDL is tuned properly. For instance, the reflectivity obtained in Fig. \ref{fig--2D_w}g, h decreases from $11.84\%$ to $2.01\%$. This highlights the importance of properly tuning the Rayleigh function in such simulations, as previously pointed by \cite{AllaertsPhD}. Conversely, the difference between the tuned RDL and the RC is more marginal, with the RDL outperforming the RC in three out of the four cases, reaching a minimum reflectivity value of $0.33\%$. Overall, simulations with $\pi_\Gamma=3.47\times 10^{-3}$ show higher reflectivity when compared to simulations with $\pi_\Gamma=3.47\times 10^{-2}$, which is again relatable to the $L_z^{ra}/\lambda_z$ ratio.

Next, we define the following error quantity
\begin{equation*}
	\mathcal{E} = \frac{\max_{x \in [0,L_x-L_x^{fr}]} \bigl\{ \bigl| w^\mathrm{Ref}(x+x_\mathrm{shift},z) - w(x,z) \bigl| \bigl\}}{\max_{x \in [0,L_x^\mathrm{Ref}-L_x^{fr}]} \bigl\{ w^\mathrm{Ref}(x,z) \bigl\}} 
\end{equation*}
where $w^\mathrm{Ref}(x,z)$ denotes the vertical velocity field obtained with the reference setup. Moreover, $x_\mathrm{shift}=90$ km so that we compare both solutions around the box-like forcing region. The idea behind this metric is to quantify the amplitude of the spurious gravity waves triggered by the fringe body force in the domain of interest with respect to the amplitude of the gravity waves triggered by the box-like force term in the reference simulation. Figure \ref{fig--2D_w} shows that $\mathcal{E}$ is almost insensitive to the upper boundary condition. However, the new fringe-region technique reduces $\mathcal{E}$ considerably. For instance, $\mathcal{E} \approx 4$ in Fig. \ref{fig--2D_w}q, r, meaning that the amplitude of the fringe-induced gravity waves is four times the maximum gravity-wave amplitude triggered by the box-like force term in the reference solution. Conversely, $\mathcal{E} =0.15$ when the wave-free fringe-region technique is applied, that is in Fig. \ref{fig--2D_w}s, t.  
\begin{figure}[t!]
	\centering
	\includegraphics[width=1\textwidth]{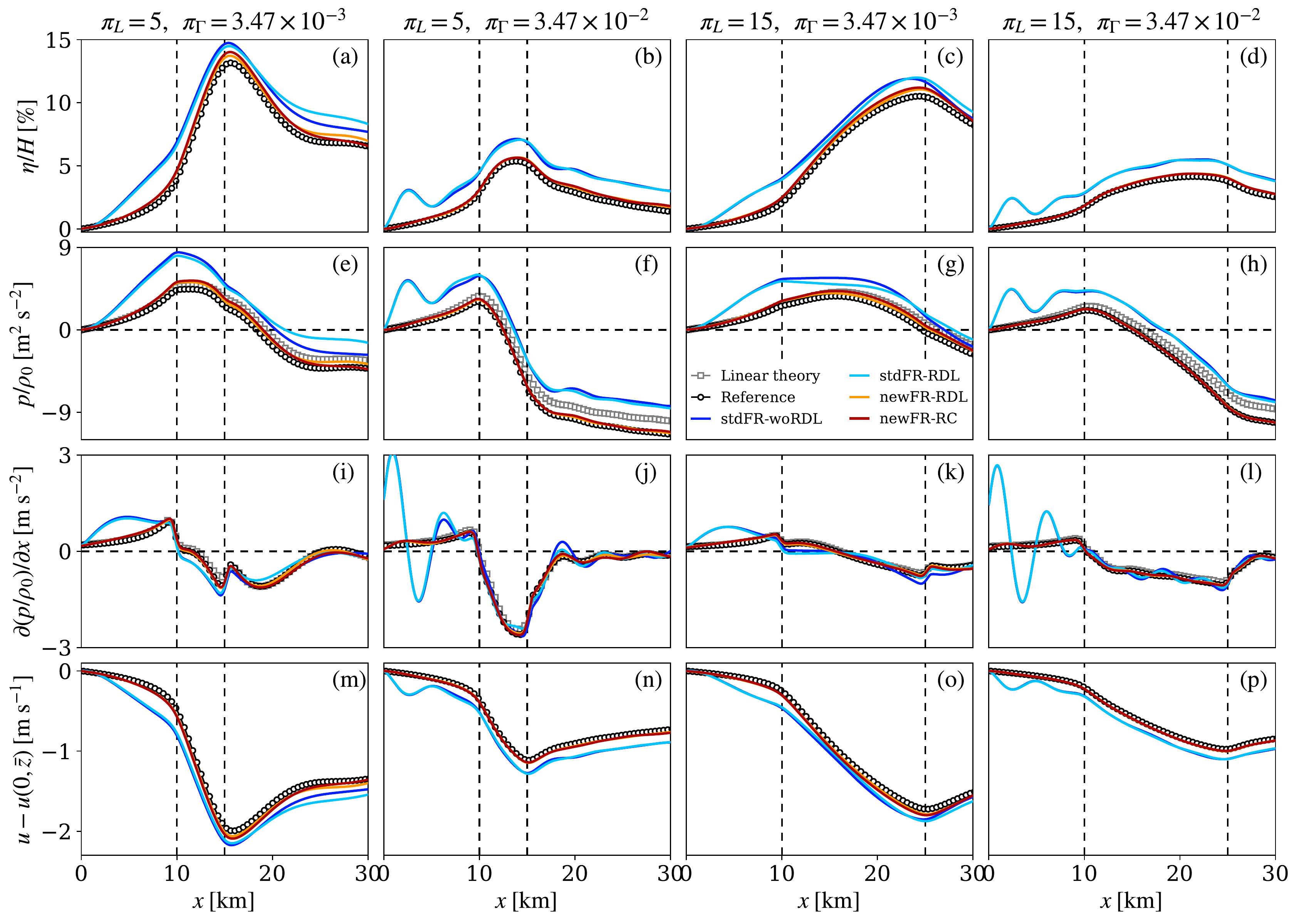}%
	\caption{(a-d) Capping inversion displacement, (e-h) pressure perturbation at $z=H$, (i-l) streamwise pressure gradient at $z=H$ and (m-p) velocity perturbation at $\bar{z}=100$ m evaluated with linear theory and with the different numerical setups. The vertical dashed lines denote the box-like forcing region}
	\label{fig--2D_all}
\end{figure}

The gravity waves shown in Fig. \ref{fig--2D_w} are triggered by the inversion layer displacement, which we denote with $\eta$. This quantity is shown in Fig. \ref{fig--2D_all}a-d for all simulation setups. Note that $\eta$ represents a streamline which is computed starting from $x=0$ and $z=H$. The reference simulation results show that a strongly stratified free atmosphere limits the vertical displacement of air parcels, therefore contributing to a lower capping inversion displacement. Very good agreement with the reference solutions is found when the wave-free fringe-region technique is applied. Conversely, the use of the standard fringe-region technique causes an overestimation of $\eta$ together with an oscillation near the inflow region. The latter is particularly visible when $\pi_\Gamma=3.47 \times 10^{-2}$.

Next, the gravity-wave induced pressure perturbations along the $x$-direction taken at $z=H$ are shown in Fig. \ref{fig--2D_all}e-h. Typical unfavourable and favourable pressure gradients are observed in front of and through the box-like forcing region, respectively \citep{Smith2010,Allaerts2017,Allaerts2017b,Allaerts2019}. However, the standard fringe-region technique overestimates the pressure build-up in the zone in front of the box-like forcing region when compared against the reference solution. This is expected, since a higher capping inversion displacement amplifies the cold temperature anomaly resulting in higher pressure perturbations within the ABL \citep{Smith2010,Allaerts2017}. Conversely, the solution obtained with the wave-free fringe-region method match well with the reference one in all cases.

Pressure perturbations can also be evaluated using linear theory. Given the Fourier coefficients of the reference simulation capping inversion displacement $\hat{\eta}^\mathrm{Ref}(k)$, the pressure Fourier coefficients given by linear theory, denoted with $\hat{p}^\mathrm{LT}(k)$, are computed as
\begin{equation*}
	\frac{1}{\rho_0}\hat{p}^\mathrm{LT}(k) = \biggl( \frac{g \Delta \theta}{\theta_0} + i \frac{N^2 - k^2 U_\infty^2}{\sqrt{N^2/U_\infty^2 - k^2}} \biggl) \hat{\eta}^\mathrm{Ref}(k)
\end{equation*}
where $i=\sqrt{-1}$ \citep{Gill1982,Nappo2002,Lin2007,Smith2010,Allaerts2019}. Figure \ref{fig--2D_all}e-h illustrates that the newFR-RDL, newFR-RC and reference pressure profiles agree well with results from linear theory in all cases. However, an offset which increases along the streamwise direction is observed. We hypothesize that this offset is caused by non-linear effects. In fact, the reference, newFR-RDL and newFR-RC profiles coincide with linear theory at $z=10$ km, where the perturbations amplitude has decreased considerably (not shown). The more accurate pressure profiles obtained with the wave-free fringe-region technique translate in more accurate pressure gradient predictions particularly in front of the box-like forcing region, as shown in Fig.~\ref{fig--2D_all}i-l.

Finally, Fig. \ref{fig--2D_all}m-p shows the velocity perturbation at $\bar{z}=100$~m. In all cases, the velocity profiles obtained with the standard fringe-region technique follow the reference solution trend within and downwind the forcing region. However, large discrepancies are observed upwind, where the profiles depart from the reference ones. Contrarily, an excellent match is observed when the wave-free fringe-region method is used. 

\subsection{Sensitivity Analysis}\label{sec--2D_sensitivity_study}
We now study how the reflectivity and $\mathcal{E}$ vary with the atmospheric state and the length of the box-like force term. To this end, we use the same numerical setups adopted in Sect. \ref{sec--2D_flow_physics}. The only difference is that we make use of a coarser grid with $N_x = 640$ to limit the computational cost. Moreover, we adopt the same $\pi_g$, $\pi_{\Delta H}$ and $\pi_L$ values of Sect. \ref{sec--2D_flow_physics} but we vary $\pi_{\Delta \theta}$ and $\pi_\Gamma$ so that the \textit{Fr} and $P_N$ values range from $0.5$ to $2.5$ for a total of $169$ atmospheric states. This allows us to extensively explore the \textit{Fr}--$P_N$ parameter space.  
\begin{figure}[t!]
	\centering
	\includegraphics[width=1\textwidth]{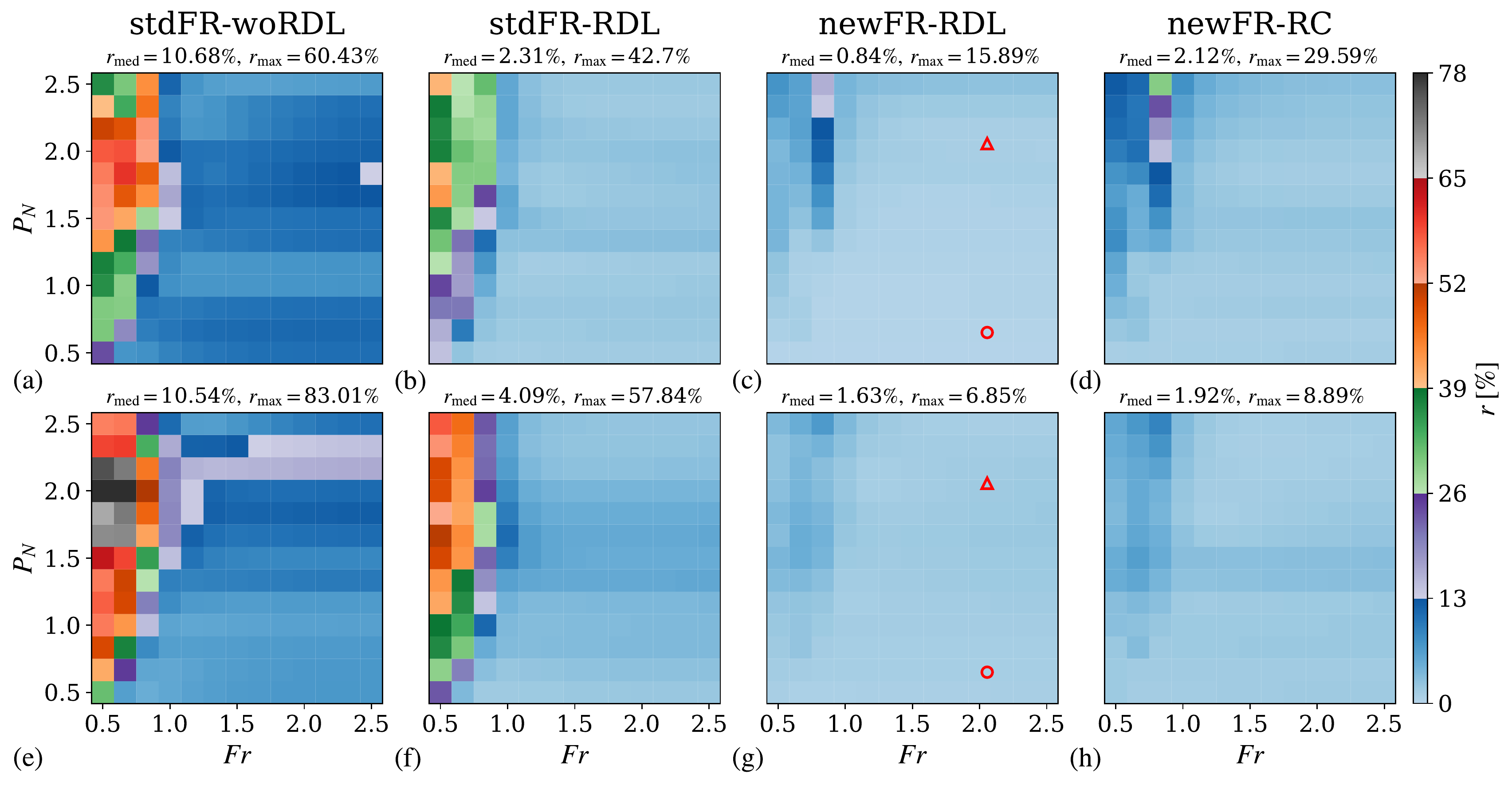}%
	\caption{Reflectivity as function of the \textit{Fr} and $P_N$ numbers obtained with (a-d) $\pi_L=5$ and (e-h) $\pi_L=15$ using the different numerical setups. The symbols $r_\mathrm{med}$ and $r_\mathrm{max}$ denote the median and the maximum of the $r$ distribution. The red triangle and circle mark the atmospheric states analyzed in details in Sect. \ref{sec--2D_flow_physics}}
	\label{fig--2D_sensitivity_r}
\end{figure}

Figure \ref{fig--2D_sensitivity_r}a-d and Fig. \ref{fig--2D_sensitivity_r}e-h show the reflectivity obtained with the different numerical setups using $\pi_L=5$ and $\pi_L=15$, respectively. The pattern observed for both $\pi_L$ values are similar. In both cases, the reflectivity is rather constant along lines of low and constant $P_N$ values while it shows a higher variability for higher $P_N$. This behaviour was also noted by \cite{Smith2010} and \cite{Lanzilao2021}, who observed that the flow response is less sensitive to changes in Froude number when $P_N$ is low. The median of the reflectivity distribution obtained with $\pi_L=15$ in the stdFR-RDL and newFR-RDL cases differs of several percentage points, going from $4.1\%$ to $1.6\%$. A more accentuate difference is observed in the maximum reflectivity value (see Fig. \ref{fig--2D_sensitivity_r}f, g). This illustrates that the absence of spurious gravity waves also reduces the amount of energy reflected downward.  In terms of upper boundary conditions, the highest reflectivity values are obtained when a RDL or a RC are not used. For instance, reflectivity values above $80\%$ are observed with the stdFR-woRDL setup. Finally, Fig.~\ref{fig--2D_sensitivity_r}c, d and Fig.~\ref{fig--2D_sensitivity_r}g, h show that the tuned RDL slightly outperforms the RC. The red triangle and circle in Fig.~\ref{fig--2D_sensitivity_r}c, g denote the simulation setup and atmospheric state used to calibrate the Rayleigh function in Sect. \ref{sec--2D_tuning_RDL}.

\begin{figure}[t!]
	\centering
	\includegraphics[width=1\textwidth]{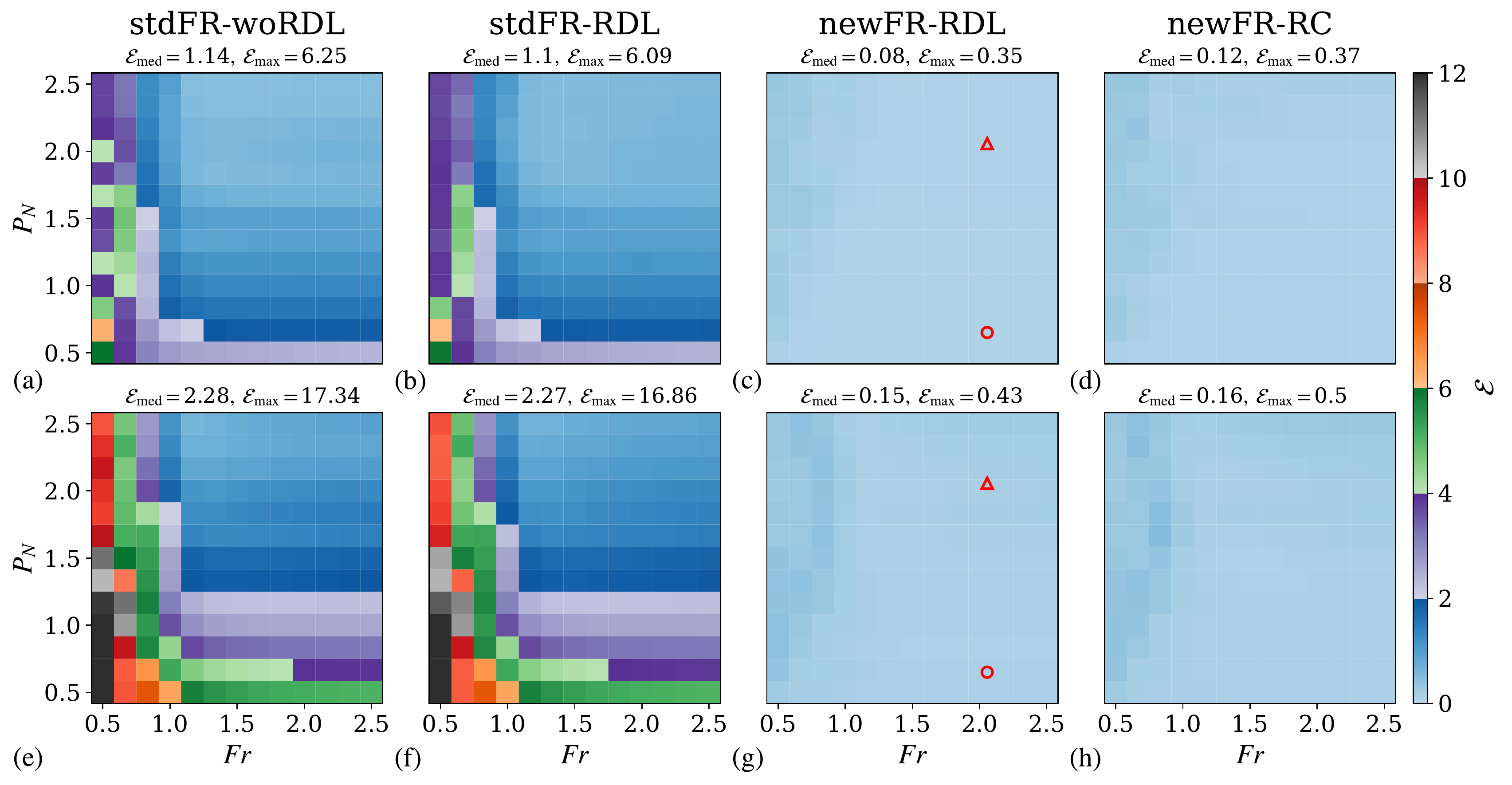}%
	\caption{$\mathcal{E}$ as function of the \textit{Fr} and $P_N$ numbers obtained with (a-d) $\pi_L=5$ and (e-h) $\pi_L=15$ using the different numerical setups. The symbols $\mathcal{E}_\mathrm{med}$ and $\mathcal{E}_\mathrm{max}$ denote the median and the maximum of the $\mathcal{E}$  distribution. The red triangle and circle mark the atmospheric states analyzed in details in Sect. \ref{sec--2D_flow_physics}}
	\label{fig--2D_sensitivity_eps}
\end{figure}

A similar analysis has been conducted using $\mathcal{E}$ as a metric. Results are shown in Fig. \ref{fig--2D_sensitivity_eps}. We observe that the upper boundary condition has limited impact on such a metric. Contrarily, a clear difference is observed between the standard and wave-free fringe-region techniques, where the median of the $\mathcal{E}$ distribution goes from $1.1$ to $0.08$ with $\pi_L=5$ and from $2.27$ to $0.15$ when $\pi_L=15$. Higher values of $\mathcal{E}$ are observed for subcritical flows and low $P_N$ numbers when the standard fringe-region technique is used, with maximum values above $17$ when $\pi_L=15$. Conversely, the wave-free fringe-region technique shows low and rather constant values of $\mathcal{E}$ across the whole $\textit{Fr}$--$P_N$ parameter space. 

The results of this sensitivity analysis point out that the wave-free fringe-region technique improves the quality of the numerical solution for all $\pi_L$, $\pi_{\Delta \theta}$ and $\pi_\Gamma$ values considered. In terms of upper boundary conditions, the RC has the advantage to be parameter-free and it does not require an additional sponge layer. Conversely, the RDL needs to be tuned and requires a buffer layer. However, the computational overhead caused by the RDL in our setup is negligible due to the limited amount of vertical grid points used within the sponge layer. Moreover, the RDL outperforms the RC in all cases analyzed. 

\section{Three-Dimensional Inviscid-Flow Simulations}\label{sec--3Dinviscidflow}
In this section we study the flow behaviour in an inviscid-flow environment with a three-dimensional computational domain. To this end, we fix the box-like forcing region length and width to $L_x^s=15$ km and $L_y^s=9$ km, respectively (see Appendix~1). This roughly corresponds to the area occupied by the London Array wind farm. The length and width of the computational domain are $L_x=100$~km and $L_y=80$~km with $N_x=1000$ and $N_y=800$, which correspond to a squared grid with resolution of $100$ m. The atmospheric states and tuning of the Rayleigh, fringe and damping functions correspond to the ones adopted in Sect. \ref{sec--2Dinviscidflow}. Moreover, results obtained with the two-dimensional inviscid-flow simulations have shown that the tuned RDL provides accurate flow fields with lower values of reflectivity when compared against the RC. Therefore, in this section we only compare numerical results obtained with the stdFR-RDL and newFR-RDL setups.

\begin{figure}[t!]
	\centering
	\includegraphics[width=1\textwidth]{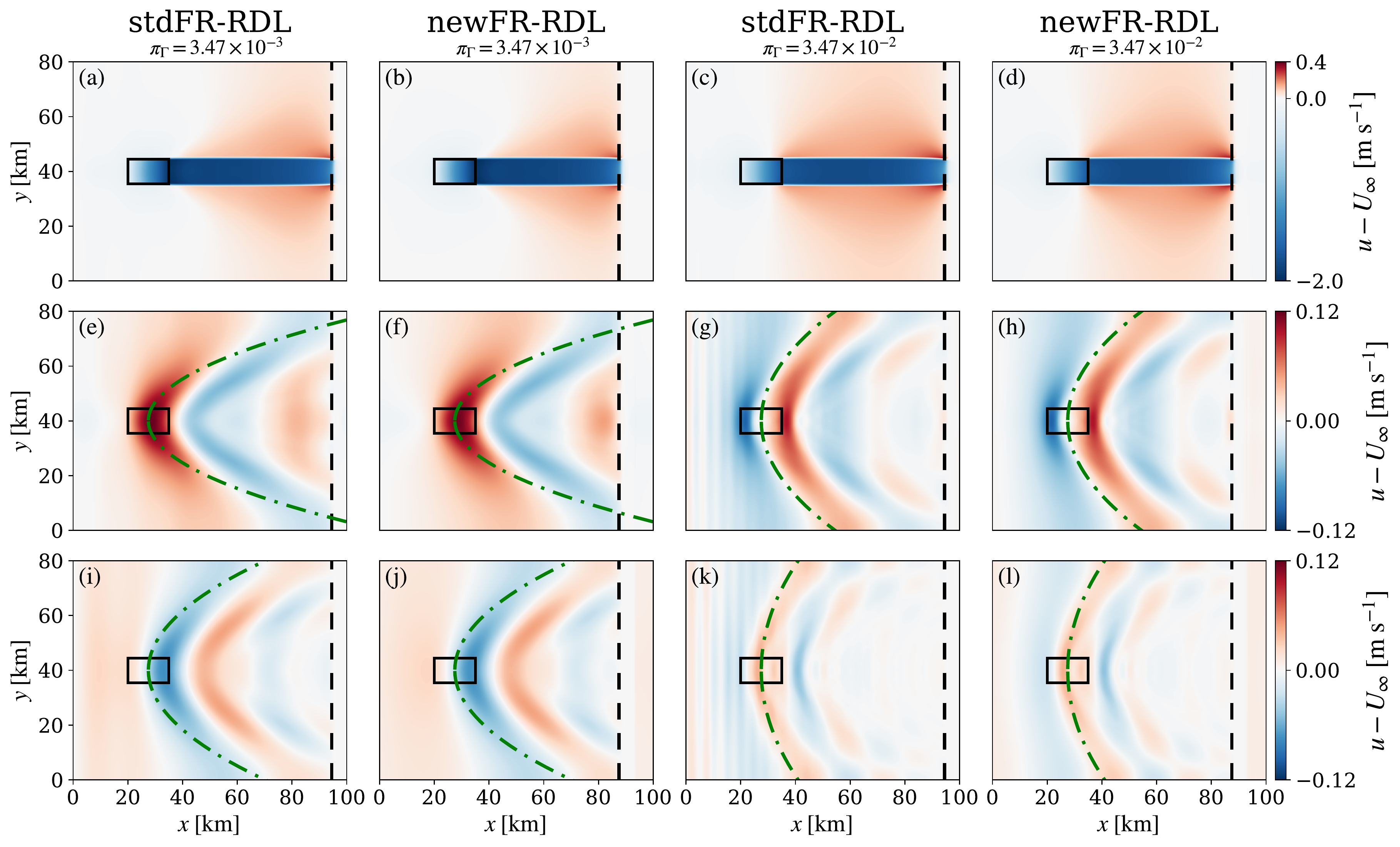}%
	\caption{Top view of streamwise velocity taken at (a-d) $z=0.1$ km, (e-h) $z=5$ km and (i-l) $z=10$ km. The black solid lines denote the region in which the box-like force term is applied. The black dashed lines indicate the starting location of the fringe region. Finally, the green dash-dotted line denotes the parabola along which wave energy is trailed downstream by internal gravity waves}
	\label{fig--3D_u}
\end{figure}
Figure \ref{fig--3D_u}a-d illustrates top views of streamwise velocity fields taken at $z=100$~m. The velocity gradually decreases over the box-like forcing region, reaching a minimum value in proximity of the trailing edge. Moreover, the wake recovery is limited due to the absence of turbulent mixing. Similar results were obtained with a two-dimensional domain (see Fig. \ref{fig--2D_all}m-p). Note that the flow is symmetric with respect to the centreline of the domain, since the Coriolis force is neglected. Finally, at the end of the domain, the fringe-region technique forces the flow to $U_\infty$, breaking the periodicity. No major differences in terms of velocity perturbations are observed at a height of $100$~m among the different numerical setups and $\pi_\Gamma$ values. We remark though that this is not anymore the case in a LES environment (see Sect. \ref{sec--LESresults}). Next, Fig. \ref{fig--3D_u}e-h shows top views of streamwise velocity taken at $z=5$ km. The stronger free lapse rate decreases the gravity-wave vertical wavelength, therefore Fig. \ref{fig--3D_u}g, h shows velocity oscillations with a lower horizontal wavelength than the ones observed in Fig. \ref{fig--3D_u}e, f. As shown previously, the standard fringe-region technique triggers spurious gravity waves which perturb the vertical velocity field. Consequently, through the continuity constraint, the streamwise velocity field is also distorted. This is mainly visible when comparing Fig. \ref{fig--3D_u}g, h in the zone upwind of the box-like forcing region. 

\cite{Smith1980} derived analytically that the internal wave energy is transported downstream along parabolas given by
\begin{equation*}
	y^2 = \frac{N a x}{U_\infty} z , \quad a = \bigl( k_x^2+ l_y^2 \bigl)^{-\frac{1}{2}}
\end{equation*}
when the flow is hydrostatic and the obstacle perturbing the flow consists of a bell-shape circular mountain. Note that $a$ represents a characteristic length scale while $k_x=1/L_x^s$ and $l_y=1/L_y^s$ \citep{Lin2007}. The dash-dotted green line in Fig. \ref{fig--3D_u}e-l represents such parabola. Despite the fact that our solution is derived for a rectangular drag force, we still find a very good agreement with the theory of \cite{Smith1980}. Finally, Fig. \ref{fig--3D_u}g, h shows top views of streamwise velocity taken at $z=10$ km. The U-shaped gravity-wave pattern widen with height, and a good agreement between theory and simulation is still found. Even at this height, the oscillations triggered by the standard fringe-region technique are still visible, mainly in Fig. \ref{fig--3D_u}k.

\section{Large-Eddy Simulations}\label{sec--LESresults}
In this section, the stdFR-RDL and newFR-RDL numerical setups are applied to an LES of a wind farm. The Coriolis force is now included in the momentum equations together with the wall stress and the sub-grid scale model, so that the full governing equations described in Sect. \ref{sec--LESmodel} are resolved. Moreover, a fully-developed turbulent inflow profile is used, which is provided by a concurrent precursor simulation. The case setup is described in Sect. \ref{sec--windfarm_setup} while results are shown in Sect. \ref{sec--windfarm_results}. In the remainder of the text, we use a bar to denote time-averaged quantities while spatial averages are denoted with angular brackets.

\subsection{Case Setup}\label{sec--windfarm_setup}
We consider a large farm with $160$ turbines disposed in $16$ rows and $10$ columns in a staggered pattern with respect to the main wind direction. The turbine-rotor diameter and hub height are $D=198$~m and $z_\mathrm{h}=119$~m, which correspond to the dimensions of the IEA $10$ MW offshore wind turbine~\citep{Bortolotti2019}. Next, the streamwise and spanwise spacings are set to $s_xD = s_yD = 5D$, which lead to a farm length and width of $14.85$ km and $9.4$~km, respectively. The box-like force term is now replaced by wind turbines, which are modelled using a non-rotating actuator disk model with a disk-based thrust coefficient of $C_\mathrm{T}'=4/3$, which corresponds to a thrust coefficient of $C_\mathrm{T}=0.75$ \citep{Calaf2010,Goit2015,Allaerts2017,Lanzilao2022}. Moreover, a simple yaw controller is implemented to keep the turbine-rotor disk perpendicular to the incident wind flow. 

The fringe and damping functions are tuned as reported in Sect. \ref{sec--newfringe}. To ensure a $10$ km distance both upwind and downwind between the farm and the fringe region, we select a domain length of $40$ km \citep{Inoue2014,Allaerts2017,Lanzilao2022}. Moreover, to minimize the sidewise blockage, we fix the width of the domain to $18$ km. The grid resolution in the streamwise and spanwise directions is set to $31.25$ m and $20$ m, respectively. With such a resolution we have a total of $10$ grid points along the turbine-rotor disk in the spanwise direction, which is similar to earlier studies \citep{Calaf2010,Allaerts2017}. In the vertical direction, we adopt the same grid used in Sects. \ref{sec--2Dinviscidflow} and \ref{sec--3Dinviscidflow}, that is a domain height of $25$ km with $490$ grid points, $10$ of which lay within the RDL. The Rayleigh function is also tuned as in Sect. \ref{sec--2D_tuning_RDL}. In fact, at the height of $15$ km, turbulence has decayed so that the velocity fields are constant with height and laminar, meaning that there is no need to re-tune the RDL. The precursor domain does not contain wind turbines nor a fringe region, therefore we select a streamwise length of $10$ km. The lateral and vertical dimensions coincide with the ones of the main domain. The combination of precursor and main simulations leads to a total of $705.6 \times 10^6$ grid cells, which is three orders of magnitude higher than the number of grid cells used in simulations performed in Sect. \ref{sec--2Dinviscidflow}. 
\begin{figure}[t!]
	\centering
	\includegraphics[width=1.0\textwidth]{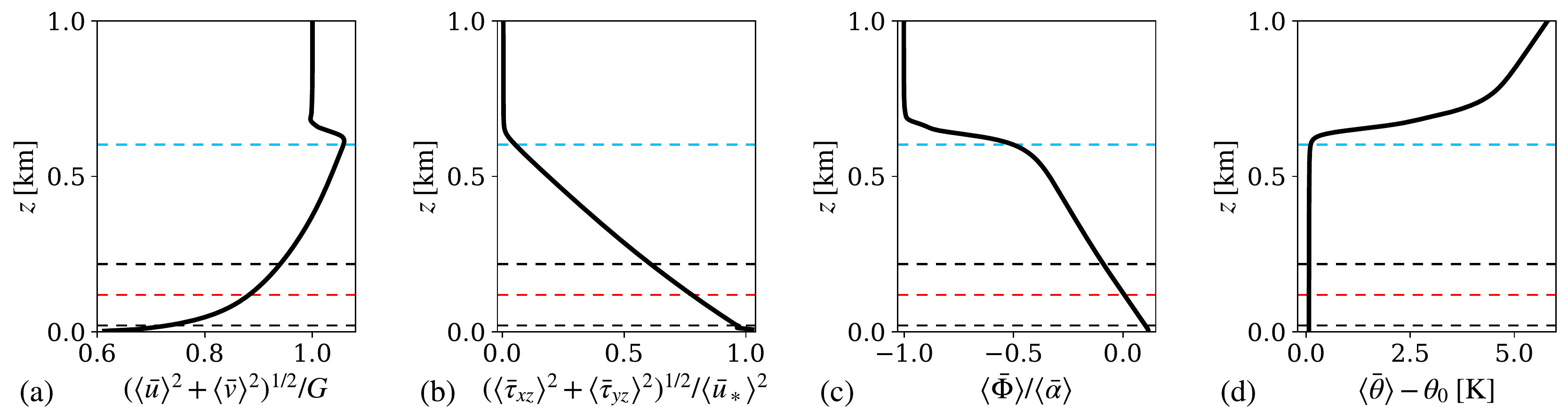}%
	\caption{Vertical profiles, averaged over the full horizontal directions and over the last two hours of the precursor simulation, of (a) horizontal velocity magnitude, (b) total shear stress magnitude, (c) horizontal wind direction and (d) potential temperature. The red dashed line denotes the hub height while the black dashed lines are representative of the rotor dimension. Finally, the blue dashed line represents the boundary-layer height}
	\label{fig--precursor}
\end{figure}

Similarly to previous sections, the atmospheric state is represented by a CNBL. The initial velocity profile in the mixed layer is provided by the \cite{Zilitinkevich1989} model with parameters $u_\ast = 0.375$~m s\textsuperscript{-1} and $z_0 = 2 \times 10^{-3}$~m, which represent the friction velocity and surface roughness, while the free atmosphere is characterized by a constant geostrophic wind $G = 10.5$~m s\textsuperscript{-1}. The two velocity profiles are then combined following the method proposed by \cite{Allaerts2015}. Next, the \cite{Rampanelli2004} model is used to specify the vertical potential temperature profile. The ground temperature is fixed to $\theta_0=282.5$~K. The base of the capping inversion is set to $H_\mathrm{in}=580$~m while the capping inversion depth and strength are fixed to $\Delta H = 175$ m and $\Delta \theta = 3.8$~K, respectively. Finally, we select a free lapse rate of $\Gamma = 5.1$ K km\textsuperscript{-1}. We note that these initial velocity and potential temperature vertical profiles corresponds to the ones adopted in \cite{Lanzilao2022}.

We further add random divergence-free perturbations with amplitude of $0.1G$ in the first $100$ m to the vertical velocity profile. This initial state is given as input to the precursor simulation which is progressed in time for $20$ h. At this point, a statistically steady state has been reached  \citep{Allaerts2016,Allaerts2017}. Figure \ref{fig--precursor} illustrates vertical profiles of several quantities averaged over the last two hours of simulation and over the horizontal directions. The boundary layer extends up to the capping inversion, located now at $H=630$~m, below which the total shear stress magnitude shows a close to linear profile. The direction of the flow at hub height is parallel to the $x$-direction (i.e. $\langle \bar{\Phi} \rangle (z_\mathrm{h}) = 0^\circ$). This is achieved by using the wind-angle controller developed by \cite{Allaerts2015}. The velocity at hub height measures $9.25$ m s\textsuperscript{-1} with a turbulence intensity of $5.35$\%. The friction velocity and the geostrophic wind angle are $\langle \bar{u}_\ast \rangle = 0.361$ m s\textsuperscript{-1} and $\langle \bar{\alpha} \rangle =-12.34^\circ$. Finally, the \textit{Fr} and $P_N$ numbers are $1.18$ and $1.12$, respectively. 

Next, we switch on the turbines in the main domain and we run main and precursor simulations in parallel. The turbulent fully-developed CNBL described above forms the inflow profile $u_{\mathrm{in},i}(\boldsymbol{x}), i=1,2,3$ which is imposed by the standard and wave-free fringe-region techniques in the main domain. This second spin-up phase lasts for $1$ h, which corresponds to approximately two wind-farm flow-through times. At this point, the boundary layer in the main domain has adapted to the additional drag force produced by the turbines and a statistically steady state has been reached again. This is suggested by the fact that negligible differences are observed in the numerical solution taken at subsequent time instances. Finally, the wind-angle controller in the precursor simulation is switched off and statistics are collected during a time window of $2$ h.
\begin{figure}[t!]
	\centering
	\includegraphics[width=1\textwidth]{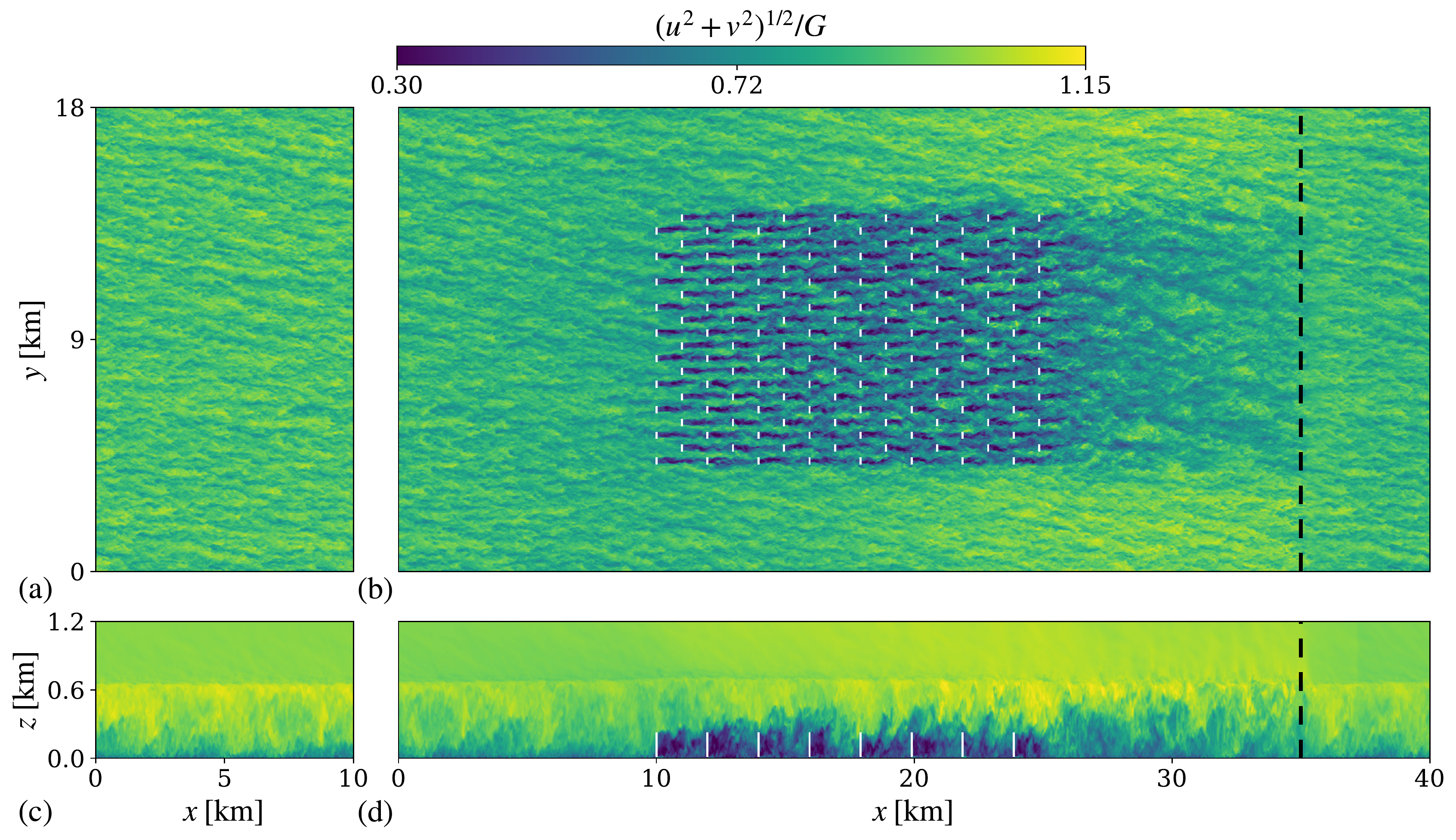}%
	\caption{Instantaneous contours of horizontal velocity magnitude at $T=2$ h obtained with the newFR-RDL setup. Top view at hub height and side view at $y=8.258$ km (i.e. along the $5^\mathrm{th}$ turbine column) of (a,c) precursor and (b,d) main simulations. The wind-turbine rotors are denoted with white lines while the vertical black dashed lines indicate the starting location of the fringe region}
	\label{fig--les_prec_main}
\end{figure}

\subsection{Results}\label{sec--windfarm_results}
A typical flow response to wind-farm forcing is shown in Fig.~\ref{fig--les_prec_main}, which displays instantaneous contours of velocity magnitude obtained at the end of the newFR-RDL simulation (i.e. $T=2$ h) for both the precursor and main domains. Figure \ref{fig--les_prec_main}a, c illustrates a top and side view of the precursor domain. The flow direction at hub height remains quasi-parallel to the $x$-direction, with  $\langle \bar{\Phi} \rangle (z_\mathrm{h}) = 0.15^\circ$. The boundary layer growth is limited by the presence of the capping inversion which shows a constant height along the streamwise direction since no momentum sinks or sources are applied within the ABL (apart from the wall stress). This precursor simulation provides the inflow condition for the main domain, which is shown in Fig.~\ref{fig--les_prec_main}b, d. In front of the farm, the wave-free fringe-region technique does not cause oscillations in the velocity magnitude. Instead, the velocity gradually decreases as a response to the gravity-wave induced unfavourable pressure gradient (see below). In the side view, we can observe the formation of an internal boundary layer which grows along the streamwise direction. However, its growth is limited by the presence of the capping inversion, which is displaced upward. Therefore, to satisfy mass conservation, the flow is redirected around the farm, as visible in Fig. \ref{fig--les_prec_main}b. For a more detailed discussion on the flow physics, we refer to \cite{Lanzilao2022}.

\begin{figure}[t!]
	\centering
	\includegraphics[width=1\textwidth]{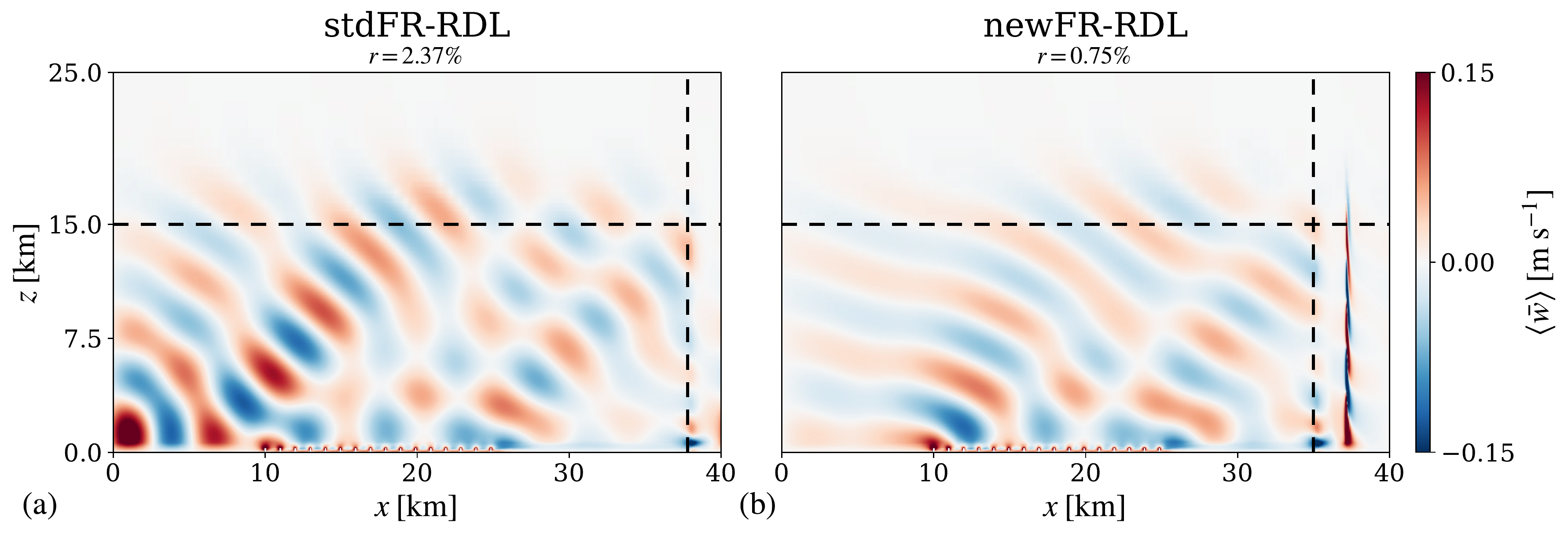}%
	\caption{Contours of vertical velocity for cases (a) stdFR-RDL and (b) newFR-RDL, averaged over time and along $x$--$z$ planes within the turbine columns footprint (i.e. along all $x$--$z$ planes that intersect the turbines rotor disk). The wind-turbine rotors are denoted with vertical white lines while the vertical and horizontal black dashed lines indicate the starting location of the fringe region and the RDL}
	\label{fig--les_w}
\end{figure}

Similarly to the inviscid-flow simulations, the upward displacement of the capping inversion caused by the farm-induced momentum sink triggers gravity waves, which are visible in Fig. \ref{fig--les_w}. Here, the time- and space-averaged vertical velocity field obtained with the stdFR-RDL and newFR-RDL setups are shown. Also here, the standard fringe-region method perturbs the flow downstream by exciting spurious gravity waves with higher amplitudes than the ones triggered by the farm. Contrarily, the wave-free fringe-region technique traps the perturbation within the fringe region, without altering the physics in the domain of interest. Moreover, the combination of the wave-free fringe-region technique with the tuned RDL leads to a reflectivity of only $0.75\%$, which is an order of magnitude lower than what was observed in previous studies \citep{Taylor2007,Wu2017,Allaerts2017,Allaerts2017b}. Finally, we notice that in both cases the flow is pushed downward at the trailing edge of the farm (i.e $x=25$~km). This negative perturbation in the vertical velocity field excites a second train of gravity waves which is out of phase with respect to the one triggered at the leading edge (i.e $x=10$~km). Whether this phenomenon occurs or not depends on the gravity-wave vertical wavelength and the length and width of the farm. Its effects on wind-farm performances is an interesting topic for future research. 
\begin{figure}[t!]
	\centering
	\includegraphics[width=1\textwidth]{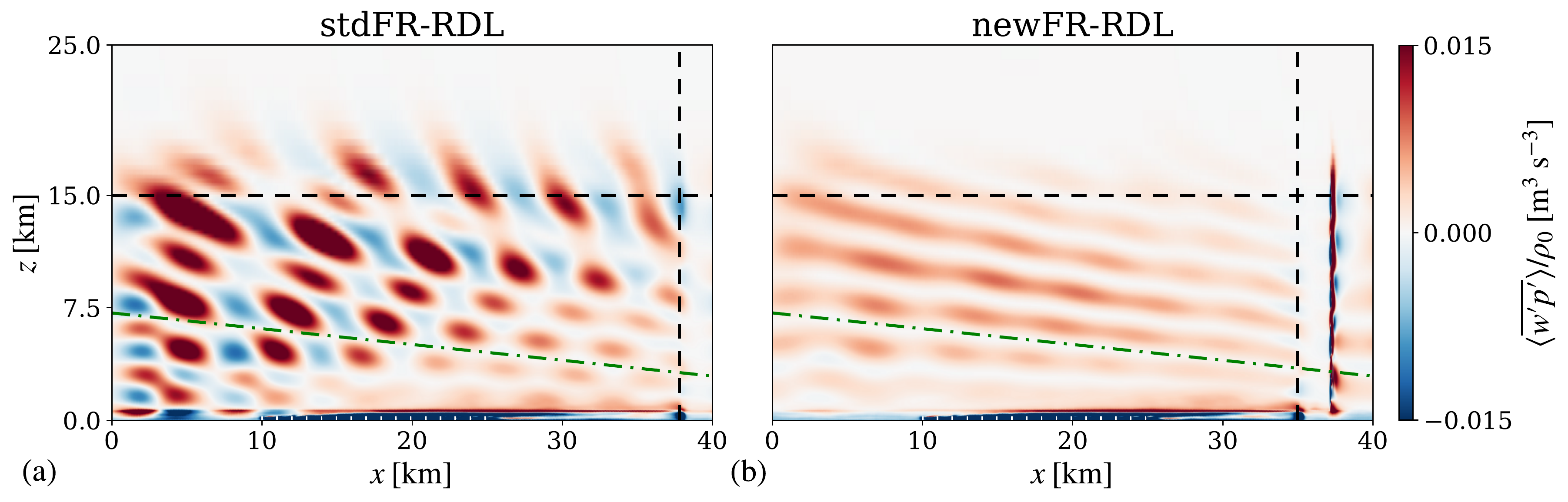}%
	\caption{(a,b) Internal wave energy flux averaged over time and along $x$–$z$ planes within the turbine columns footprint obtained with the stdFR-RDL and newFR-RDL setups, respectively. The wind-turbine rotors are denoted with vertical white lines while the vertical and horizontal black dashed lines indicate the starting location of the fringe region and the RDL. Finally, the green dash-dotted line denotes the inclination of the gravity-wave phase line predicted by linear theory}
	\label{fig--les_wp}
\end{figure}
\begin{figure}[t!]
	\centering
	\includegraphics[width=1\textwidth]{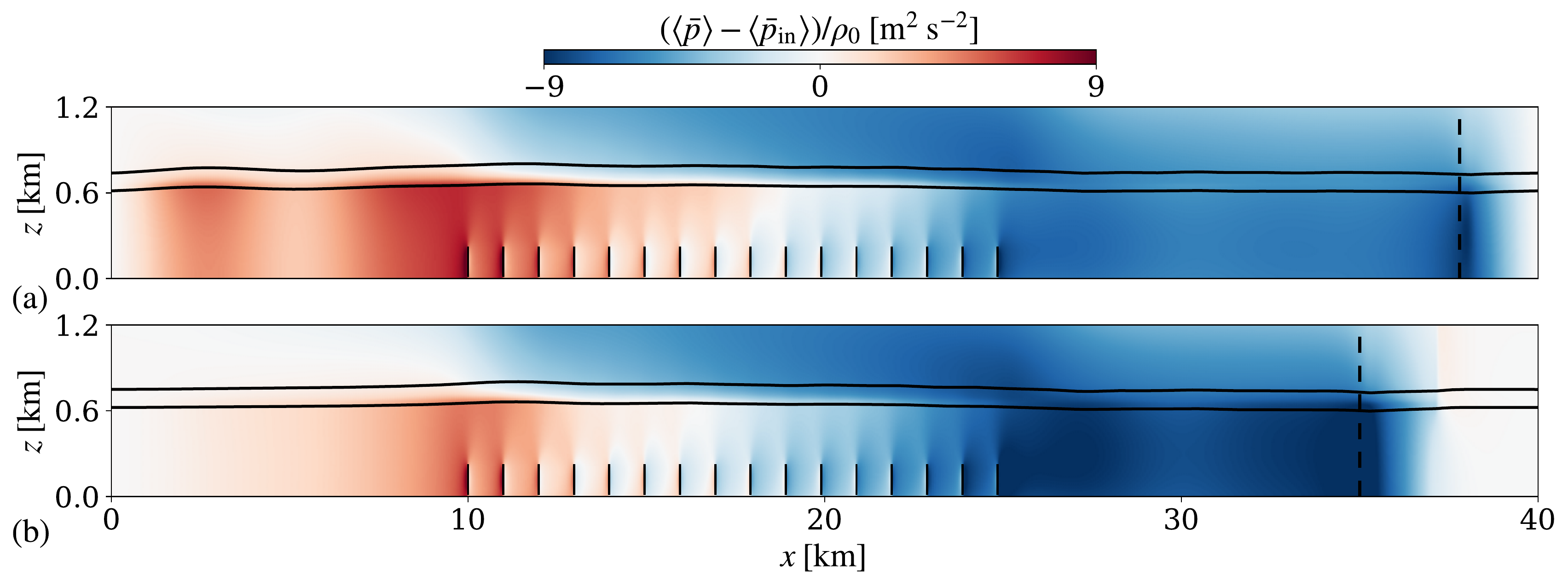}%
	\caption{Contours of pressure perturbation with respect to the pressure measured at the inflow for cases (a) stdFR-RDL and (b) newFR-RDL, averaged over time and along $x$--$z$ planes within the turbine columns footprint. The black solid lines show the time-averaged evolution of the inversion-layer base and top while the wind-turbine rotors are denoted with vertical black lines. The vertical black dashed lines indicate the starting location of the fringe region}
	\label{fig--les_p}
\end{figure}

The internal gravity waves shown in Fig. \ref{fig--les_w} transport energy upward. This is illustrated in Fig. \ref{fig--les_wp}a, b, which displays a side view of the time- and space-averaged internal wave energy flux obtained with the standard and new fringe-region technique, respectively. Spurious gravity waves increase the amount of energy transported upward (red patches). Consequently, more energy is also reflected downward (blue patches). This gives rise to the checkerboard patter visible in Fig. \ref{fig--les_wp}a. Contrarily, Fig. \ref{fig--les_wp}b shows that energy is primarily transported upward and along lines parallel to the wave-front, which is in agreement with gravity-wave theory \citep{Nappo2002,Lin2007}. Moreover, the dash-dotted green line, which represents the angle that the gravity-wave phase lines make with the horizontal calculated using linear theory, is parallel to the actual phase lines predicted by our LES solver. 

The time- and space-averaged pressure perturbation within the first $1.2$~km obtained with the stdFR-RDL and newFR-RDL setups are shown in Fig. \ref{fig--les_p}a, b, respectively. A favourable and unfavourable pressure gradient in the induction region of the farm is observed in Fig. \ref{fig--les_p}a, which in turn causes a flow slow-down and speed-up (not shown). This non-physical behaviour is provoked by the spurious gravity waves induced by the standard fringe forcing. In fact, the velocity decreases monotonically in the farm-induction region when the wave-free fringe-region technique is used (not shown). Moreover, the pressure build-up for the stdFR-RDL simulation at the first turbine row is roughly $35\%$ higher than in the newFR-RDL case. The solid black lines in Fig. \ref{fig--les_p} denote the bottom and top of the capping inversion. The spurious gravity waves triggered by the standard fringe-region technique cause a vertical displacement of the capping inversion near the inflow which is comparable to the one caused by the wind farm (see Fig.~\ref{fig--les_p}a). As expected, local maxima in the pressure distribution correspond to local maxima of the inversion layer displacement, and vice versa. 
\begin{figure}[t!]
	\centering
	\includegraphics[width=1\textwidth]{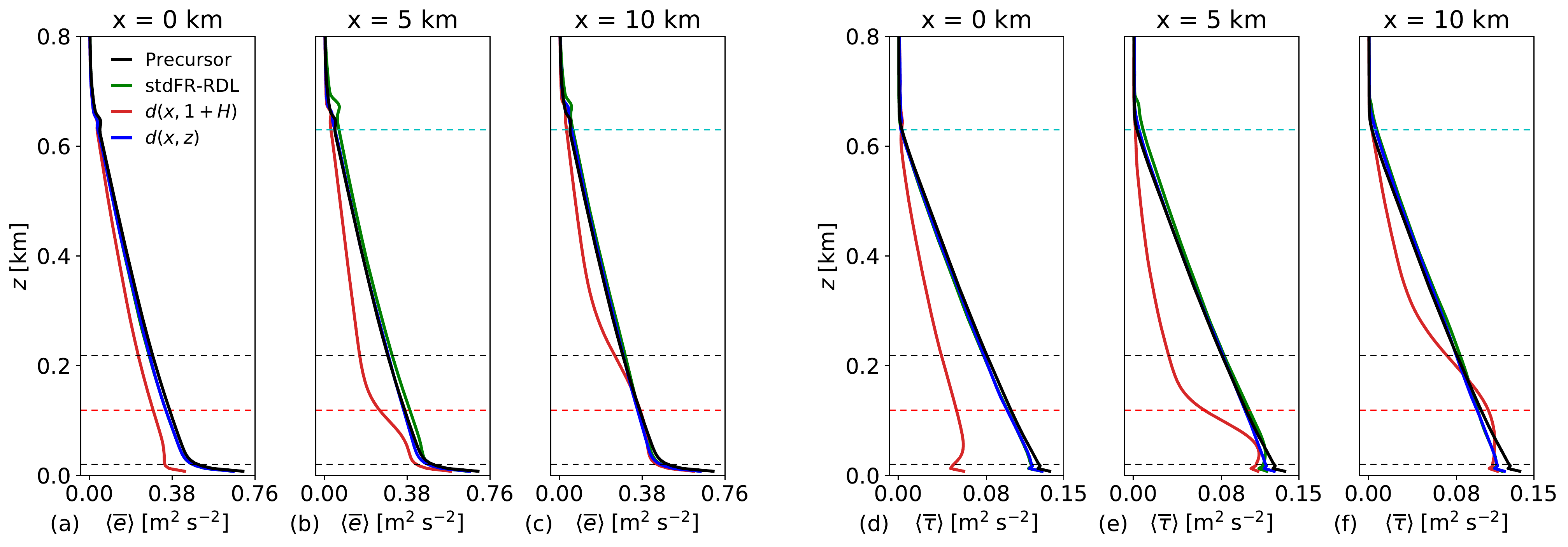}%
	\caption{(a-c) Total (i.e. resolved and sub-grid scale contribution) turbulent kinetic energy $e$ and (d-f) total shear stress magnitude $\tau$ averaged over time and over the full spanwise direction obtained at the main-domain inflow ($x=0$ km), $x=5$ km and at the location of the first-row turbines ($x=10$ km). The green line represents the results obtained with the stdFR-RDL setup while the red and blue lines denote the results obtained with the newFR-RDL setups using a height-independent (i.e. $d(x,1+H)$) and height-dependent (i.e. $d(x,z)$) damping function, respectively. Moreover, the red dashed line denotes the hub height while the black dashed lines are representative of the rotor dimension. Finally, the light blue dashed line represents the boundary-layer height evaluated in the precursor domain}
	\label{fig--les_second_order_moments}
\end{figure}

Finally, Fig. \ref{fig--les_second_order_moments} shows a comparison of total turbulent kinetic energy $e$ and total shear stress magnitude $\tau$ obtained using the standard and wave-free fringe-region technique. For the latter, we show two cases. The first one adopts a height-dependent damping function (i.e. Eq. \ref{eq--damping_function}) while the second one uses $d(x,1+H)$ which implies that the damping is also active inside the ABL (since $\mathcal{H}(1)=1$). It is evident from Fig. \ref{fig--les_second_order_moments} that the latter case not only dampen spurious gravity waves but also turbulent fluctuations. In fact, both $e$ and $\tau$ are underestimated at the main-domain inflow (i.e. $x=0$ km). However, turbulence develops in the main domain so that the turbulence characteristics converge toward the precursor ones for increasing streamwise location. Contrarily, when the convective term in the vertical momentum equation is only dampened above the ABL, the turbulence characteristics observed at the main-domain inflow are comparable with the precursor ones. A similar behaviour is observed for the standard fringe-region technique. This suggests that, in case of a turbulent inflow profile, the wave-free fringe-region technique should adopt a $x$- and $z$-dependent damping function.

\section{Conclusions}\label{sec--conclusions}
The aim of the current study was to develop a numerical setup which imposes the inflow and upper boundary conditions without distorting the numerical solution in LES of wind farms operating in stratified atmospheres. To this end, we implemented and compared two non-reflective upper boundary conditions, that is the RDL and the RC. Furthermore, we noticed that the standard fringe-region technique triggers spurious gravity waves. Therefore, we have developed a wave-free fringe-region technique which imposes the inflow condition while limiting spurious effects on the surrounding flow. We achieved this by locally damping the convective term in the vertical momentum equation. The different numerical setups have been tested in both inviscid- and viscid-flow environments.

We started our analysis with inviscid-flow simulations on a two-dimensional domain ($x$--$z$). First, we proposed a method to properly calibrate the Rayleigh function to minimize wave reflection together with the computational burden that comes with this additional sponge layer. Results showed that a tuned RDL can reduce the reflectivity of several percentage points if compared with simulations that do not have such a layer. Moreover, the RDL outperformed the RC in all cases analyzed. Next, we have shown that the standard fringe forcing triggers spurious gravity waves. The amplitude of such waves can be $3$ to $4$ times higher than the amplitude of gravity waves induced by the smooth box-like force term. This mainly causes an overestimation of the pressure build-up in front of the forcing region together with non-physical streamwise velocity oscillations. Conversely, the solution obtained with the wave-free fringe-region technique closely follows the reference one. Good agreement with gravity-wave linear theory was also observed. Next, the different numerical setups have been applied to $169$ atmospheric states and two different box-like forcing region lengths. Reflectivity values up to $80\%$ were observed in subcritical flow conditions when the standard fringe-region technique was used. Moreover, in all cases, the new fringe-region technique outperformed the standard method, imposing the inflow conditions with a minimal impact on the surrounding flow. Overall, the tuned RDL in combination with the wave-free fringe-region technique (i.e. newFR-RDL) was the setup which provided the best performance, with reflectivity values below $1\%$ in most of the cases.

Further, we extended the analysis to a three-dimensional domain, where the box-like forcing region emulated the momentum sink generated by a large wind farm. Also in this case, the newFR-RDL setup allowed us to impose the inflow condition without altering the flow fields downstream. Moreover, a good agreement between our numerical solutions and the theory of \cite{Smith1980} was observed.

Finally, we have compared the performance of the standard and wave-free fringe-region techniques in an LES of a large wind farm operating in a CNBL. Despite the use of a turbulent inflow condition, we observed similar results. The standard fringe-region technique introduced spurious gravity waves in the domain of interest. These waves caused an oscillation in the pressure distribution, which led to a $30\%$ higher pressure build-up in front of the farm when compared against the results obtained with the wave-free fringe-region method. In fact, in the latter case, the pressure and the streamwise velocity components showed monotonic trends in the farm induction region. We note that a reflectivity of only $0.75\%$ was obtained with the newFR-RDL setup, which is an order of magnitude lower than what was observed in previous studies. 

In light of these results, we conclude that a properly tuned Rayleigh function in combination with a wave-free fringe-region technique provide an effective framework for LES of wind farms that operates in stratified atmospheres. In the future, we are planning to adopt this numerical setup to further study gravity-wave effects on wind-farm operations.

\begin{acknowledgements}
The authors acknowledge support from the Research Foundation Flanders (FWO, grant no. G0B1518N), and from the project FREEWIND, funded by the Energy Transition Fund of the Belgian Federal Public Service for Economy, SMEs, and Energy (FOD Economie, K.M.O., Middenstand en Energie). The computational resources and services in this work were provided by the VSC (Flemish Supercomputer Center), funded by the Research Foundation Flanders (FWO) and the Flemish Government department EWI.
\end{acknowledgements}

\noindent\small{\textbf{Data Availability}} The datasets generated and/or analysed during the current study are available from the corresponding author on reasonable request.
\normalsize

\section*{Appendix 1: Smooth Box-Like Force Model}\label{app--force_model}
In the inviscid-flow simulations, we use a smooth box-like force model which emulates the presence of a wind-farm, generating a momentum sink within the ABL. Such a force term is defined by the function

\begin{equation}
f_1(\boldsymbol{x}) = \beta \frac{U_\infty^2}{D_s} \frac{\mathcal{S}(x,L_x^s,\delta_x,x_0) \mathcal{S}(y,L_y^s,\delta_y,y_0) \mathcal{S}(z>0,2L_z^s,\delta_z,0)}{s_c(L_x^s,\delta_x) s_c(L_y^s,\delta_y) s_c(L_z^s,\delta_z/2)} 
\label{eq--box_forcing}
\end{equation}
where 
\begin{equation*}
\small
\mathcal{S}(s,L,\delta,s_0) = 
\begin{cases}
	\cos{\bigl( \pi \bigl[s-(s_0 - L/2 + \delta)/4 \delta\bigl]}\bigl),& \text{if } s_0 - L/2-\delta < s < s_0 - L/2 + \delta\\
	1,& \text{if } s_0 - L/2+\delta < s < s_0 + L/2 - \delta\\
	\cos{\bigl( \pi \bigl[s-(s_0 + L/2 - \delta)/4 \delta\bigl]}\bigl),& \text{if } s_0 + L/2-\delta < s <  s_0 + L/2 + \delta\\
	0,              & \text{otherwise}
\end{cases}
\end{equation*}
\begin{equation*}
\small
s_c(L,\delta) = L+2\delta (4-\pi)/\pi.
\end{equation*}
The parameters $L_x^s+2\delta_x$, $L_y^s+2\delta_y$ and $L_z^s+\delta_z$ determine the size of the box-like forcing region, with $\delta_x$, $\delta_y$ and $\delta_z$ controlling the smoothness of the profiles to minimize the role of non-linear effects. Moreover, the scaling term $s_c$ ensures that the integrated force over the whole computational domain equals to $\beta U_\infty^2/D_s$, where $U_\infty$ denotes the inflow velocity, $D_s$ represents a length scale related to the turbine-rotor diameter while $\beta$ is a non-dimensional parameter which regulates the magnitude of the force term. Following \cite{AllaertsPhD}, we fix $U_\infty=12$ m s\textsuperscript{-1}, $D_s=100$ m and $\beta=0.01$. Finally, the centre of the forcing region along the $x$- and $y$-direction is denoted with $x_0$ and $y_0$. Note that the momentum sink is applied only to the $u$-momentum equation, meaning that $f_2(\boldsymbol{x}) =f_3(\boldsymbol{x}) =0$. 

In the two-dimensional inviscid-flow simulations, the box-like forcing region starts at $100$ km when the reference setup is used and at $10$ km in the other cases. Here, we use two $L_x^s$ values, that is $5$ km and $15$ km, with $L_z^f=0.6$~km. The smoothing parameters along the $x$- and $z$-direction are fixed to $\delta_x=0.5$~km and $\delta_z=0.4$~km. We note that $H= L_z^s + \delta_z$, meaning that the force is applied within the whole region below the capping inversion, similarly to \cite{Smith2010,Smith2022}. The same setup is used in the three-dimensional inviscid-flow simulations, with the difference that $L_x^s=15$ km, $L_y^s=9$ km and $\delta_y=0.5$~km. Note that in the two-dimensional simulations we fix $\mathcal{S}(y,L_y^s,\delta_y,y_0)=1$ and $s_c(L_y^s,\delta_y)=1$.

\section*{Appendix 2: Grid Sensitivity Study}\label{app--grid_sensitivity}
\begin{figure}[t!]
\centering
\includegraphics[width=1\textwidth]{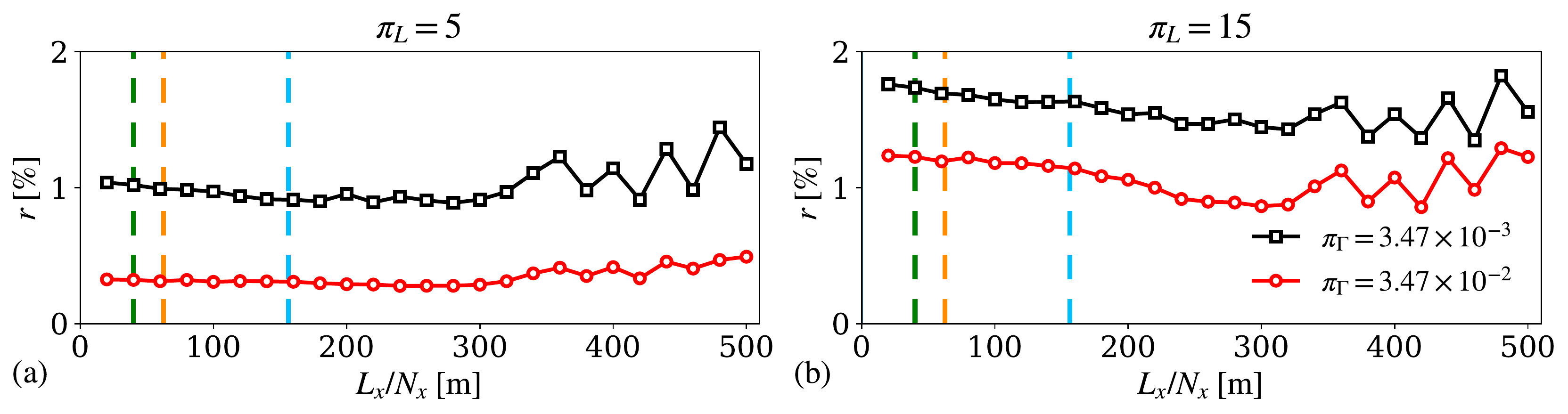}%
\caption{Grid sensitivity analysis for (a) $\pi_L=5$ and (b) $\pi_L=15$ performed with the newFR-RDL numerical setup (hence $L_x=40$ km). The light blue, orange and green vertical dashed lines refer to the grid resolution adopted in Sects. \ref{sec--2D_tuning_RDL}, \ref{sec--2D_sensitivity_study} and \ref{sec--2D_flow_physics}, respectively}
\label{fig--grid_sensitivity}
\end{figure}

We perform a grid sensitivity analysis to determine the dependence of the reflectivity on the grid resolution in an inviscid-flow environment. To this end, we use the newFR-RDL setup described in Section \ref{sec--2D_flow_physics} and we vary the grid resolution in the $x$-direction spanning from $500$ m to $20$ m with $L_x=40$~km. Results obtained with the atmospheric states adopted in Sect. \ref{sec--2D_flow_physics} are shown in Fig. \ref{fig--grid_sensitivity}. When $\pi_L=5$ and $\pi_\Gamma=3.47 \times 10^{-2}$, the reflectivity obtained on a grid with resolution of $156.25$ m (used in Sect. \ref{sec--2D_tuning_RDL}) and $40$ m (used in Sect. \ref{sec--2D_flow_physics}) is $0.32\%$ and $0.31\%$, respectively. A similar difference is observed when a weaker free lapse rate is adopted (i.e $\pi_\Gamma=3.47 \times 10^{-3}$), although some oscillations in the $r$ value take place at coarser grid resolutions. An analogous pattern is shown in Fig. \ref{fig--grid_sensitivity}b. These results suggest that the numerical solution is grid independent, justifying the changes in grid resolution made throughout Sects.~\ref{sec--2Dinviscidflow} and \ref{sec--3Dinviscidflow}.

\begin{figure}[t!]
\centering
\includegraphics[width=1\textwidth]{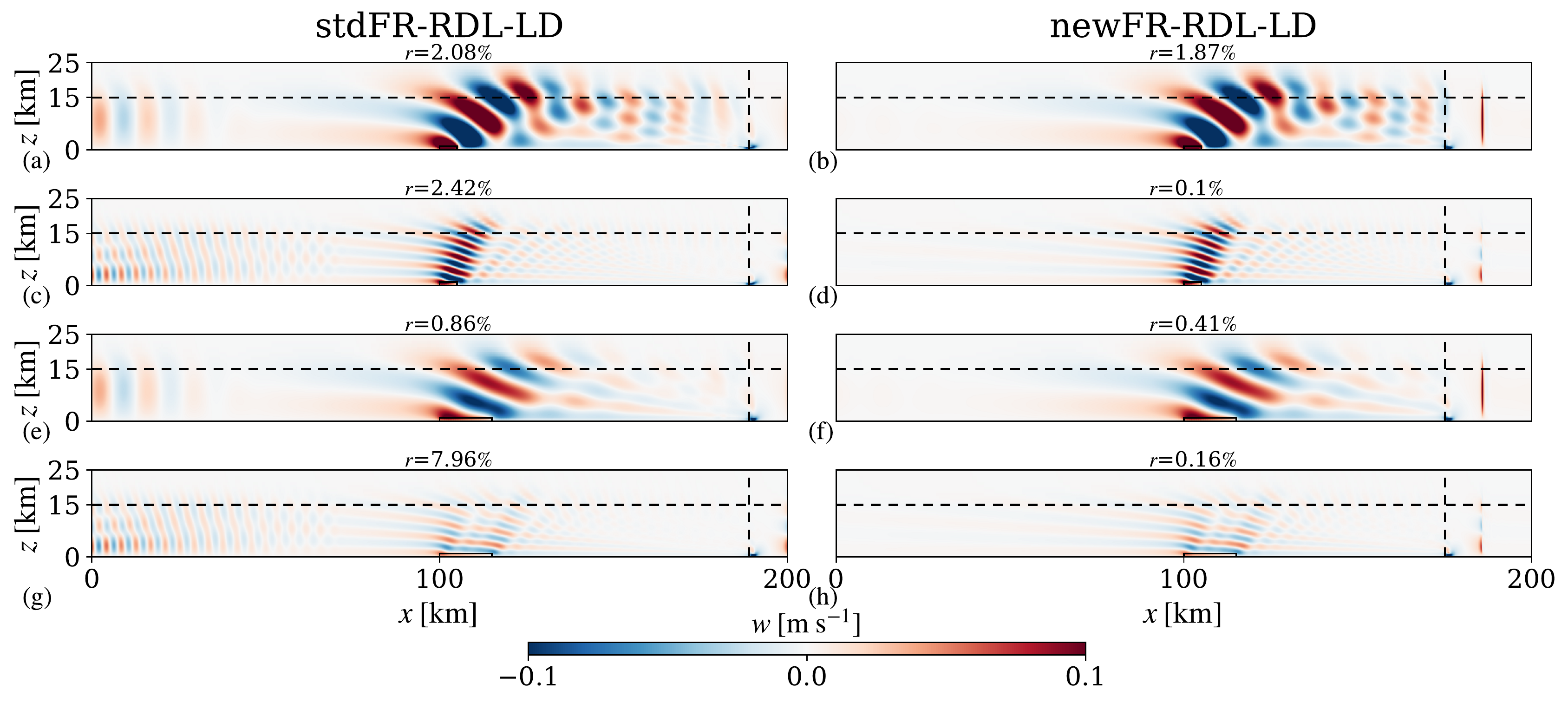}%
\caption{Side view of vertical velocity obtained with (a,b) $\pi_L=5$ and $\pi_\Gamma=3.47 \times 10^{-3}$, (c,d) $\pi_L=5$ and $\pi_\Gamma=3.47 \times 10^{-2}$, (e,f) $\pi_L=15$ and $\pi_\Gamma=3.47 \times 10^{-3}$ and (g,h) $\pi_L=15$ and $\pi_\Gamma=3.47 \times 10^{-2}$. The snapshots display the full-domain reference solution obtained with the standard and wave-free fringe-region techniques. The black solid lines denote the region in which the box-like force term is applied. The black horizontal and vertical dashed lines indicate the starting location of the RDL and the fringe region}
\label{fig--w_sideview_ref}
\end{figure}

\section*{Appendix 3: Two-Dimensional Inviscid-Flow Simulations Computed on the Long Domain}\label{app--ref_sim}
In this appendix, we display the full-domain vertical velocity field obtained using the stdFR-RDL-LD and newFR-RDL-LD (also named as Reference) setups defined in Sect. \ref{sec--2D_flow_physics}. Figure~\ref{fig--w_sideview_ref} illustrates the vertical velocity obtained with different $\pi_L$ and $\pi_\Gamma$ values. With a domain length of $200$ km, it is even more evident how the standard fringe-region technique triggers spurious gravity waves, perturbing the flow field downstream. With such a long domain, the perturbations die out before reaching the box-like forcing region. Therefore, the phase lines of the gravity waves induced by the momentum sink are not distorted. Contrarily, the vertical velocity fields obtained with the wave-free fringe-region method do not alter the flow physics in the domain of interest. Moreover, Fig. \ref{fig--w_sideview_ref}b, d, f, h also shows that the domain length is large enough for the newFR-RDL-LD solution to be considered as the reference, since it is not affected by the boundary conditions. Finally, the wave-free fringe-region technique also reduces the reflectivity. For instance, we measure reflectivity values of $7.96\%$ and $0.16\%$ with the standard and wave-free fringe-region method, respectively, when $\pi_L=15$ and $\pi_\Gamma=3.47\times 10^{-2}$ (see Fig.~\ref{fig--w_sideview_ref}g, h).


\begin{thebibliography}{56}
	\providecommand{\natexlab}[1]{#1}
	\providecommand{\url}[1]{{#1}}
	\providecommand{\urlprefix}{URL }
	\expandafter\ifx\csname urlstyle\endcsname\relax
	\providecommand{\doi}[1]{DOI~\discretionary{}{}{}#1}\else
	\providecommand{\doi}{DOI~\discretionary{}{}{}\begingroup
		\urlstyle{rm}\Url}\fi
	\providecommand{\eprint}[2][]{\url{#2}}
	
	\bibitem[{Allaerts(2016)}]{AllaertsPhD}
	Allaerts D (2016) Large-eddy simulation of wind farms in conventionally neutral
	and stable atmospheric boundary layers. PhD thesis
	
	\bibitem[{Allaerts and Meyers(2015)}]{Allaerts2015}
	Allaerts D, Meyers J (2015) Large eddy simulation of a large wind-turbine array
	in a conventionally neutral atmospheric boundary layer. Physics of Fluids
	27(6):065,108, \doi{10.1063/1.4922339}
	
	\bibitem[{Allaerts and Meyers(2016)}]{Allaerts2016}
	Allaerts D, Meyers J (2016) Effect of inversion-layer height and coriolis
	forces on developing wind-farm boundary layers. AIAA SciTech 6:521--538
	
	\bibitem[{Allaerts and Meyers(2017)}]{Allaerts2017}
	Allaerts D, Meyers J (2017) Boundary-layer development and gravity waves in
	conventionally neutral wind farms. J Fluid Mech 814:95--130
	
	\bibitem[{Allaerts and Meyers(2018)}]{Allaerts2017b}
	Allaerts D, Meyers J (2018) Gravity waves and wind-farm efficiency in neutral
	and stable conditions. Boundary-Layer Meteorol 166:269--299
	
	\bibitem[{Allaerts and Meyers(2019)}]{Allaerts2019}
	Allaerts D, Meyers J (2019) Sensitivity and feedback of wind-farm induced
	gravity waves. J Fluid Mech 862:990 -- 1028
	
	\bibitem[{Bennett(1976)}]{Bennett1976}
	Bennett AF (1976) Open boundary conditions for dispersive waves. J Atmos Sci
	33:176--182
	
	\bibitem[{Bortolotti et~al.(2019)Bortolotti, Tarrés, Dykes, Merz, Sethuraman,
		Verelst, and Zahle}]{Bortolotti2019}
	Bortolotti P, Tarrés HC, Dykes K, Merz K, Sethuraman L, Verelst D, Zahle F
	(2019) Iea wind task 37 on systems engineering in wind energy wp2.1 reference
	wind turbines. Technical report
	
	\bibitem[{Bougeault(1982)}]{Bougeault1982}
	Bougeault P (1982) A non-reflective upper boundary condition for limited-height
	hydrostatic models. Monthly Weather Review 111:420--429
	
	\bibitem[{Béland and Warn(1975)}]{Beland1975}
	Béland M, Warn T (1975) The radiation condition for transient rossby waves. J
	Atmos Sci 32:1873--1880
	
	\bibitem[{Calaf et~al.(2010)Calaf, Meneveau, and Meyers}]{Calaf2010}
	Calaf M, Meneveau C, Meyers J (2010) Large eddy simulation study of fully
	developed wind-turbine array boundary layers. Phys Fluids 22:015,110
	
	\bibitem[{Devesse et~al.(2022)Devesse, Lanzilao, Jamaer, van Lipzig, and
		Meyers}]{Devesse2022}
	Devesse K, Lanzilao L, Jamaer S, van Lipzig N, Meyers J (2022) Extending the
	applicability of a wind-farm gravity-wave model to vertically non-uniform
	atmospheres. Wind Energy Science Discussions 2022:1--25,
	\doi{10.5194/wes-2021-138}
	
	\bibitem[{Dhamankar et~al.(2015)Dhamankar, Blaisdell, and
		Lyrintzis}]{Dhamankar2015}
	Dhamankar NS, Blaisdell GA, Lyrintzis AS (2015) Overview of turbulent inflow
	boundary conditions for large-eddy simulations. Aiaa Journal 56(4):1317--1334
	
	\bibitem[{Doyle et~al.(2005)Doyle, Shapiro, Jiang, and Bartels}]{Doyle2005}
	Doyle JD, Shapiro MA, Jiang Q, Bartels DL (2005) Large-amplitude mountain wave
	breaking over greenland. J Atmos Sci 62:3106--3126
	
	\bibitem[{Durran and Klemp(1983)}]{Durran1983}
	Durran DR, Klemp JB (1983) A compressible model for the simulation of moist
	mountain waves. Mon Wea Rev 111:2341--2361
	
	\bibitem[{Gadde and Stevens(2021)}]{Gadde2021}
	Gadde SN, Stevens RJAM (2021) Interaction between low-level jets and wind farms
	in a stable atmospheric boundary layer. Phys Rev Fluids 6:014,603,
	\doi{10.1103/PhysRevFluids.6.014603}
	
	\bibitem[{Gill(1982)}]{Gill1982}
	Gill AE (1982) Atmosphere-ocean dynamics. International Geophysics Series
	30:Academic Press, San Diego, USA
	
	\bibitem[{Goit and Meyers(2015)}]{Goit2015}
	Goit JP, Meyers J (2015) Optimal control of energy extraction in wind-farm
	boundary layers. J Fluid Mech 768:5--50
	
	\bibitem[{Hu(2008)}]{Hu2008a}
	Hu FQ (2008) Development of pml absorbing boundary conditions for computational
	aeroacoustics: A progress review. Computers and Fluids 37:336--348
	
	\bibitem[{Hu et~al.(2008)Hu, Li, and Lin}]{Hu2008b}
	Hu FQ, Li XD, Lin DK (2008) Absorbing boundary conditions for nonlinear euler
	and navier–stokes equations based on the perfectly matched layer technique.
	Journal of Computational Physics 227:4398--4424
	
	\bibitem[{Inoue et~al.(2014)Inoue, Matheou, and Teixeira}]{Inoue2014}
	Inoue M, Matheou G, Teixeira J (2014) Les of a spatially developing atmospheric
	boundary layer: Application of a fringe method for the stratocumulus to
	shallow cumulus cloud transition. Monthly Weather Review 142:1365–1393
	
	\bibitem[{Jiang and Doyle(2004)}]{Jiang2004}
	Jiang Q, Doyle JD (2004) Gravity wave breaking over the central alps: Role of
	complex terrain. J Atmos Sci 61:2249--2266
	
	\bibitem[{Klemp and Durran(1982)}]{Klemp1982}
	Klemp JB, Durran DR (1982) An upper boundary condition permitting internal
	gravity wave radiation in numerical mesoscale models. Monthly Weather Review
	111:430--444
	
	\bibitem[{Klemp and Lilly(1977)}]{Klemp1977}
	Klemp JB, Lilly DK (1977) Numerical simulations of hydrostatic mountain waves.
	Journal of the atmospheric sciences 35:78--107
	
	\bibitem[{Klemp et~al.(2018)Klemp, Dudhia, and Hassiotis}]{Klemp2008}
	Klemp JB, Dudhia J, Hassiotis AD (2018) An upper gravity-wave absorbing layer
	for nwp applications. Monthly Weather Review 136:3987--4004
	
	\bibitem[{Lanzilao and Meyers(2021)}]{Lanzilao2021}
	Lanzilao L, Meyers J (2021) Set-point optimization in wind farms to mitigate
	effects of flow blockage induced by atmospheric gravity waves. Wind Energy
	Science 6(1):247--271
	
	\bibitem[{Lanzilao and Meyers(2022)}]{Lanzilao2022}
	Lanzilao L, Meyers J (2022) Effects of self-induced gravity waves on finite
	wind-farm operations using a large-eddy simulation framework. Journal of
	Physics: Conference Series 2265(2):022,043,
	\doi{10.1088/1742-6596/2265/2/022043}
	
	\bibitem[{Lin(2007)}]{Lin2007}
	Lin YL (2007) Mesoscale Dynamics. Cambridge University Press,
	\doi{10.1017/CBO9780511619649}
	
	\bibitem[{Lundbladh et~al.(1999)Lundbladh, Berlin, Skote, Hildings, Choi, Kim,
		and Henningson}]{Lundbladh1999}
	Lundbladh A, Berlin S, Skote M, Hildings C, Choi J, Kim J, Henningson DS (1999)
	An efficient spectral method for a simulation of incompressible flow over a
	flat plate. Trita-mek Tech Rep: 11 KTH
	
	\bibitem[{Maas and Raasch(2022)}]{Maas2022}
	Maas O, Raasch S (2022) Wake properties and power output of very large wind
	farms for different meteorological conditions and turbine spacings: a
	large-eddy simulation case study for the german bight. Wind Energy Science
	7(2):715--739, \doi{10.5194/wes-7-715-2022}
	
	\bibitem[{Mason and Thomson(1992)}]{Mason1992}
	Mason PJ, Thomson DJ (1992) Stochastic backscatter in large-eddy simulations of
	boundary layers. Journal of Fluid Mechanics 242:51–78,
	\doi{10.1017/S0022112092002271}
	
	\bibitem[{Moeng(1984)}]{Moeng1984}
	Moeng CH (1984) A large-eddy-simulation model for the study of the planetary
	boundary-layer turbulence. Journal of the atmospheric science 41:2052–2062
	
	\bibitem[{Munters and Meyers(2018)}]{Munters2018}
	Munters W, Meyers J (2018) Dynamic strategies for yaw and induction control of
	wind farms based on large-eddy simulation and optimization. Energies 11:177
	
	\bibitem[{Munters et~al.(2016{\natexlab{a}})Munters, Meneveau, and
		Meyers}]{Munters2016}
	Munters W, Meneveau C, Meyers J (2016{\natexlab{a}}) Shifted periodic boundary
	conditions for simulations of wall-bounded turbulent flows. Physics of Fluids
	28(2):025,112, \doi{10.1063/1.4941912}
	
	\bibitem[{Munters et~al.(2016{\natexlab{b}})Munters, Meneveau, and
		Meyers}]{Munters2016b}
	Munters W, Meneveau C, Meyers J (2016{\natexlab{b}}) Turbulent inflow precursor
	method with time-varying direction for large-eddy simulations and
	applications to wind farms. Boundary-layer meteorology 159(2):305--328
	
	\bibitem[{Nappo(2002)}]{Nappo2002}
	Nappo CJ (2002) An introduction to atmospheric gravity waves. International
	Geophysics Series 85:Academic Press, Waltham, USA
	
	\bibitem[{Nordstrom et~al.(1999)Nordstrom, Nordin, and
		Henningson}]{Nordstrom1999}
	Nordstrom J, Nordin N, Henningson D (1999) The fringe region technique and the
	fourier method used in the direct numerical simulation of spatially evolving
	viscous flows. SIAM J Sci Comput 20:1365–1393
	
	\bibitem[{Parrish and Hu(2009)}]{Hu2009}
	Parrish SA, Hu FQ (2009) Pml absorbing boundary conditions for the linearized
	and nonlinear euler equations in the case of oblique mean flow. Int J Numer
	Meth Fluids 60:565--589
	
	\bibitem[{Porté-Agel et~al.(2020)Porté-Agel, Bastankhah, and
		Shamsoddin}]{PorteAgel2020}
	Porté-Agel F, Bastankhah M, Shamsoddin S (2020) Wind-turbine and wind-farm
	flows: A review. Boundary-Layer Meteorol 174:1--59
	
	\bibitem[{Powers et~al.(2017)Powers, Klemp, Skamarock, Davis, and
		Dudhia}]{Powers2017}
	Powers JG, Klemp JB, Skamarock WC, Davis CA, Dudhia J (2017) The weather
	research and forecasting model: overview, system efforts, and future
	directions. Bulletin of the American Meteorological Society 98
	
	\bibitem[{Rampanelli and Zardi(2004)}]{Rampanelli2004}
	Rampanelli G, Zardi D (2004) A method to determine the capping inversion of the
	convective boundary layer. J Appl Meteor 43:925--933
	
	\bibitem[{Schlatter et~al.(2005)Schlatter, Adams, and Kleiser}]{Schlatter2005}
	Schlatter P, Adams N, Kleiser L (2005) A windowing method for periodic
	inflow/outflow boundary treatment of non-periodic flows. Journal of
	Computational Physics 206:505--535
	
	\bibitem[{Smedman et~al.(1997)Smedman, Bergstrom, and Grisogono}]{Smedman1997}
	Smedman A, Bergstrom H, Grisogono B (1997) Evolution of stable internal
	boundary layers over a cold sea. Journal of Geophysical Research
	102:1091--1099
	
	\bibitem[{Smith(1980)}]{Smith1980}
	Smith RB (1980) Linear theory of stratified hydrostatic flow past an isolated
	mountain. Tellus 32:348--364
	
	\bibitem[{Smith(2010)}]{Smith2010}
	Smith RB (2010) Gravity wave effects on wind farm efficiency. Wind Energy
	13:449--458
	
	\bibitem[{Smith(2022)}]{Smith2022}
	Smith RB (2022) A linear theory of wind farm efficiency and interaction.
	Journal of the Atmospheric Sciences \doi{10.1175/JAS-D-22-0009.1}
	
	\bibitem[{Spalart and Watmuff(1993)}]{Spalart1993}
	Spalart PR, Watmuff JH (1993) Experimental and numerical study of a turbulent
	boundary layer with pressure gradients. J Fluid Mech 249:337--371
	
	\bibitem[{Stevens et~al.(2000)Stevens, Moeng, and Sullivan}]{Stevens2000}
	Stevens B, Moeng CH, Sullivan PP (2000) Entrainment and subgrid length scales
	in large-eddy simulations of atmospheric boundary-layer flows. Symposium on
	Developments in Geophysical Turbulence 58:253--269
	
	\bibitem[{Stevens et~al.(2014)Stevens, Graham, and Meneveau}]{Stevens2014}
	Stevens RJAM, Graham J, Meneveau C (2014) A concurrent precursor inflow method
	for large eddy simulations and applications to finite length wind farms.
	Renewable Energy 68:46--50
	
	\bibitem[{Stieren et~al.(2021)Stieren, Gadde, and Stevens}]{Stieren2021}
	Stieren A, Gadde SN, Stevens RJAM (2021) Modeling dynamic wind direction
	changes in large eddy simulations of wind farms. Renewable Energy
	170:1342--1352
	
	\bibitem[{Sutherland(2010)}]{Sutherland2010}
	Sutherland BR (2010) Internal gravity waves. Cambridge University Press
	
	\bibitem[{Taylor and Sarkar(2007)}]{Taylor2007}
	Taylor JR, Sarkar S (2007) Internal gravity waves generated by a turbulent
	bottom ekman layer. J Fluid Mech 590:331--354
	
	\bibitem[{Taylor and Sarkar(2008)}]{Taylor2008}
	Taylor JR, Sarkar S (2008) Direct and large eddy simulations of a bottom ekman
	layer under an external stratification. Int J Heat Fluid Flow 29:721--732
	
	\bibitem[{Teixeira(2014)}]{Teixeira2014}
	Teixeira MAC (2014) The physics of orographic gravity wave drag. Frontier in
	Physics 2
	
	\bibitem[{Verstappen and Veldman(2003)}]{Verstappen2003}
	Verstappen RWCP, Veldman AEP (2003) Symmetry-preserving discretization of
	turbulent flow. Journal of Computational Physics 187:343–368
	
	\bibitem[{Wu and Porté-Agel(2017)}]{Wu2017}
	Wu KL, Porté-Agel F (2017) Flow adjustment inside and around large finite-size
	wind farms. Energies 10:2164
	
	\bibitem[{Zilitinkevich(1989)}]{Zilitinkevich1989}
	Zilitinkevich SS (1989) Velocity profiles, the resistance law and the
	dissipation rate of mean flow kinetic energy in a neutrally and stably
	stratified planetary boundary layer. Boundary-Layer Meteorol 46:367–387
	
\end{thebibliography}
\end{document}